\documentstyle[preprint,tighten,eqsecnum,aps,prd,amssymb,amsbsy,epsf]{revtex}
\draft 
\title{Aperture synthesis for gravitational-wave data analysis: Deterministic 
Sources}  
\author{Lee Samuel Finn} 
\address{Center for Gravitational Physics and 
Astronomy\thanks{Also Department of Physics and Department of Astronomy 
and Astrophysics; e-mail {\tt LSF5@PSU.Edu}.},
The Pennsylvania State University, University Park PA 
16802}
\date{\today}
\begin{document}
\maketitle
\begin{abstract}   
Gravitational wave detectors now under construction are sensitive to
the phase of the incident gravitational waves.  Correspondingly, the
signals from the different detectors can be combined, in the analysis,
to simulate a single detector of greater amplitude and directional
sensitivity: in short, aperture synthesis.  Here we consider
the problem of aperture synthesis in the special case of a search for
a source whose waveform is known in detail: \textit{e.g.,} compact
binary inspiral.  We derive the likelihood function for joint output
of several detectors as a function of the parameters that describe the
signal and find the optimal matched filter for the detection of the
known signal.  Our results allow for the presence of noise that is
correlated between the several detectors.  While their derivation is
specialized to the case of Gaussian noise we show that the results
obtained are, in fact, appropriate in a well-defined,
information-theoretic sense even when the noise is non-Gaussian in
character.

The analysis described here stands in distinction to ``coincidence
analyses'', wherein the data from each of several detectors is studied
in isolation to produce a list of candidate events, which are then
compared to search for coincidences that might indicate common origin
in a gravitational wave signal. We compare these two analyses --- 
optimal filtering and coincidence --- in a series of numerical 
examples, showing that the optimal filtering analysis always yields a 
greater detection efficiency for given false alarm rate, even when the 
detector noise is strongly non-Gaussian. 
\end{abstract}


\section{Introduction}\label{sec:intro}

Several large interferometric gravitational wave detectors
\cite{luck00a,coles00a,marion00a} will soon be operating, perhaps to
be joined by several as or more sensitive acoustic resonant detectors
\cite{conti00a,dewaard00a,blair00a} in the years to come.  There are
several reasons for wanting to combine, in some way the, the data from
these detectors in order to draw inferences about the presence of a
signal or the parameters that characterize it: for example,
observations of a signal with several detectors increase the degree of
confidence in the detection and characterization of a signal, and the
use of several geographically separated detectors can allow one to
disentangle source parameters (e.g., sky position, polarization) that
are degenerate when observed in a single detector.

The signal response of both interferometric and acoustic detectors is
sensitive to the phase of the incident gravitational waves;
consequently, we have the opportunity to combine interfere the
response of several detectors, synthesizing a single, effectively
larger and more directionally sensitive detector: in short, aperture
synthesis.  Here we describe a likelihood-based analysis for the
detection of a signal of known, parameterized form in the joint output
of a network of detectors.  Such an analysis makes the most effective
use of the information available from each detector, in exactly the
same sense that optimal filter described in
\cite{finn92a,finn93a,cutler94a} is best-suited for detecting or
characterizing radiation incident on a single detector.

The analysis described here stands in contrast to a ``coincidence''
analysis, whereby ``events'' are identified independently in separate
detectors and these independent lists of events are later brought
together and compared for coincidences
\cite{jaranowski94a,jaranowski96a}.  The key difference between the
two analyses is their sensitivity to inter-detector correlations.  The
response of a network of gravitational wave detectors to an incident
plane wave is phase coherent.  This phase coherence is captured in the
likelihood function describing the output of a network of detectors;
however, it is absent in a coincidence analysis.  This leads the
likelihood analysis to have an increased detection efficiency for a
given false alarm rate when compared to a coincidence analysis
searching for the same signals and based on the same detector output.

We demonstrate this directly in a set of three numerical examples,
based on a network consisting of two separated detectors.  In each
example we search for a signal of known waveform arriving at an
arbitrary time and from an arbitrary direction. Two different search 
algorithms are investigated. The first is based on the maximum of the 
likelihood of the joint detector output; the second is based on the 
analysis of coincidences between the optimally filtered output of each 
detector, considered separately. The three examples differ in the 
character of the simulated noise: in the examples the noise is either 
Normal, strongly leptokurtic or strongly platykurtic. In all cases the 
likelihood analysis described here gives a substantially larger 
detection efficiency for a fixed false alarm rate. 

Most previous work on multi-detector gravitational wave data analysis
has focused on coincidence analyses or variations on coincidence
analyses \cite{jaranowski94a,jaranowski96a,creighton99a} that occur
separately at each detector.  The work described here parallels
recent, independent work by Bose and collaborators
\cite{bose99a,bose00a}.  It goes beyond that work by allowing from the
beginning detectors that are not coincident and accommodating the
possibility that the noise between pairs of detectors may be
correlated.

Section \ref{sec:nomenclature} introduces terminology and nomenclature
used throughout this work.  In section \ref{sec:gauss} the likelihood
function for the joint output of a set of gravitational wave detectors
is derived.  The form derived here is exact when the detector noise is
Normal; however, we show that it is also the best choice (in the sense
of making the least additional assumptions about the noise statistics)
when only the detector cross- and auto-correlations are known.  In
section \ref{sec:examples} we describe and present the results of
several numerical examples, meant to compare likelihood and
coincidence-based approaches to data analysis.  Finally, in section
\ref{sec:conclusions} we summarize our conclusions.

\section{Nomenclature}\label{sec:nomenclature}

\subsection{Continuous and discrete time signals}

While gravitational wave detectors are analog devices, the detector
output, including diagnostic channels containing physical and
environmental monitors, will be quantized in magnitude and sampled
discretely in time.  Data analysis will operate exclusively with this
discrete time series.

Properly handled, quantization of the continuous amplitude contributes
a small, white, additive component to the detector noise \cite[Chapter
3.7.3]{oppenheim89a}, which we will not discuss further here. 

When certain other conditions hold, the continuous in time and
discretely sampled time series are equivalent and fully
interchangeable.  In this paper we have occasion to refer to both discrete
and continuous time representations of the same underlying process.
To distinguish between these two representations, we refer to the
continuous time representation of the process $x$ by $x(t)$, where $t$
denotes continuous time, and the discrete time representation of the
same process by $x[k]$, where integer $k$ denotes the sample number.

\subsection{Scalar-, vector- and matrix-valued functions and
sequences} 

For our purposes, the output of a single gravitational-wave detector
is a real, scalar-valued time series.  An important purpose of this
paper is to discuss data analysis for a gravitational-wave
\textit{receiver,} by which we mean a logical collection of detectors. 
The output of a receiver is a vector-valued time series.  The elements
of the vector at any given sample are the detector outputs at the
corresponding moment of time.

In this paper we will have occasion to refer to scalar-, vector- and
matrix-valued time series.  To distinguish between these different
cases, we use a lowercase italic face (as in $a$) to denote scalars or
scalar-valued sequences, a lowercase bold-italic face (as in
$\bbox{a}$) to denote vectors or vector sequences, and an uppercase
bold-italic face (as in $\bbox{A}$) to denote matrices or
matrix-valued sequences (functions).

\subsection{Discrete Fourier transform}

Many different conventions for the discrete Fourier transform (DFT) 
can be found in the literature.  We adopt the conventions described in 
this section.  If you compare the notation used here with other 
conventions found in the literature, it will be to your advantage to 
pay careful attention to the normalization, index range of 
the input and output sequences, and sign convention of the transform 
kernel.

Suppose that $x[k]$ is a sequence of coefficients of length $N$, with 
initial index $k_0$.  The DFT $\widetilde{x}[k]$ of the sequence 
$x[k]$ is the periodic sequence given by
\begin{mathletters}
\begin{equation}
\widetilde{x}[j] = \sum_{k=k_0}^{k_0+N-1}x[k]\omega_N^{jk},
\end{equation}
where 
\begin{equation}
\omega_N \equiv \exp(-2\pi i/N)
\end{equation}
\end{mathletters}
is the $N^{\text{th}}$ root of unity. (Note our use of the 
engineering convention for the transformation kernel.)

The kernel of the transform, $\omega_N^{jk}$, satisfies an orthogonality
relationship,
\begin{equation}
\delta_{kl} = 
{1\over N}\sum_{n=0}^{N-1}
\omega_N^{kn}\omega_N^{-nl}
\end{equation}
as $k$ and $l$ run from $0$ to $N-1$; consequently, the DFT is
invertible. The inverse DFT (IDFT) is given by
\begin{equation}
  \check{x}[n] = {1\over
    N}\sum_{k=0}^{N-1}\widetilde{x}[k]\omega_{N}^{-nk}.
  \label{eq:invDFT}
\end{equation}
Note that the IDFT is also a periodic sequence. In particular, if
$\tilde{x}[k]$ is the DFT of a sequence $x[j]$, then $\check{x}[k]$ is
the periodic extension of $x[j]$.

A wealth of useful results and insights into the DFT/IDFT pair are
found in \cite{briggs95a}; additionally, \cite{oppenheim89a} discusses
the use of the DFT/IDFT pair in digital signal processing.

\subsection{Terminology}

A \textit{receiver} is a collection of one or more individual
gravitational-wave \textit{detectors} whose output is analyzed
collectively to draw a single set of inferences regarding the presence
and properties of radiation sources. The elements of the receiver are
the detectors, which need not be co-located, co-oriented, or of
identical design, sensitivity or bandwidth. For example, a single
receiver might consist of 
\begin{itemize}
\item the three LIGO interferometers (a 4~Km and a 2~Km interferometer
  in Washington State, USA, and a 4~Km interferometer interferometer
  in Louisiana, USA) \cite{coles00a},
\item the Virgo interferometer (located in Pisa, Italy \cite{marion00a}), 
\item the GEO-600 interferometer (located in Hannover, Germany 
\cite{luck00a}), 
\item the TAMA-300 interferometer (located in Tokyo, Japan 
\cite{tsubono95a}), 
\item the ALLEGRO detector (a resonant mass detector located in Louisiana 
\cite{hamilton97a}), 
\item the AURIGA detector (a resonant mass detector located in Padua 
\cite{cerdonio97a}), 
\item the Explorer detector (a resonant mass detector located in
  Frascati \cite{astone93a}), and
\item the Nautilus detector (a resonant mass detector located in
  Geneva \cite{astone97b}).
\end{itemize}
A gravitational-wave \textit{receiver} is thus a logical grouping of
gravitational-wave \textit{detectors.}

The output of detector $\alpha$ is the scalar-valued time-series
$g_{\alpha}$. The output $\bbox{g}$ of a receiver composed of several
detectors $\alpha,\ldots,\omega$ is the direct sum of the output of the
individual detectors:
\begin{equation}
\bbox{g} = g_{\alpha}\oplus\cdots\oplus g_{\omega}.
\end{equation}

The receiver response $\bbox{m}$ to an incident gravitational-wave 
$\bbox{h}$ depends on parameters $\bbox{\theta}$ that reflect both the 
intrinsic properties of the source and the relationship of the source 
to the receiver (i.e., distance and orientation).  For example, 
if the source is a stochastic gravitational-wave signal, then 
$\bbox{\theta}$ describes the signal spectrum and anisotropy; 
alternatively, if the source is an inspiraling compact binary system, 
then $\bbox{\theta}$ describes the binary's component masses, 
spins and orbital characteristics.  In this paper I discuss analysis 
for deterministic signals: i.e., signals like those that arise 
from binary inspiral or rapidly rotating non-axisymmetric neutron 
stars, whose waveform can be accurately described in terms of a small 
number of parameters that characterize the source and its orientation 
with respect to the detector.

\section{The Likelihood Function}
\label{sec:gauss}

\subsection{Introduction}\label{sec:likeIntro}

Let $H_0$ and $H_{\bbox{\theta}}$ denote alternative, exclusive
hypotheses regarding the presence or absence of a signal:
\begin{mathletters}
\begin{eqnarray}
H_{0} &=& \left(\begin{tabular}{l}
proposition that the receiver\\
output consists only of noise
\end{tabular}
\right),\label{eq:H0}\\
H_{\bbox{\theta}} &=& \left(
\begin{tabular}{l}
{proposition that the receiver}\\
{output consists of signal $\bbox{s}_{\bbox{\theta}}$}\\
{superposed with receiver noise}
\end{tabular}
\right)\label{eq:Htheta}
\end{eqnarray}
\end{mathletters}
where $\bbox{\theta}$ represents a set of parameters that
differentiate signals in the detector: for example, if the signal is
from an inspiraling binary neutron star or black hole system, the
parameters include the component masses and the source distance from
and orientation with respect to the receiver. In this section we
derive the likelihood function $\Lambda(\bbox{g}|\bbox{\theta})$,
describing the probability that $\bbox{g}$ is observed under the
hypothesis $H_{\bbox{\theta}}$.

Why the likelihood function?  Recall that the likelihood is any
function that, viewed as a function of $\bbox{\theta}$, is
proportional to $P(\bbox{g}|H_{\bbox{\theta}})$, the probability of
observing $\bbox{g}$ given the fixed hypothesis $H_{\bbox{\theta}}$. 
In a Bayesian analysis, the goal is to determine or characterize
$P(H_{\bbox{\theta}}|\bbox{g})$, the probability of the hypothesis
given the observation.  Through Bayes Theorem, this quantity is
directly related to the likelihood.  In frequentist statistical
analysis the goal is to determine the improbability of observing
$\bbox{g}$ under the alternative hypotheses that the signal is absent
or that it is present, or to establish confidence intervals --- a
range of hypotheses $H_{\bbox{\theta}}$ for which the observation has,
in a certain sense, high probability \cite{feldman98a}.  In either
case the likelihood plays a central role and no other function of the
observation and hypotheses offers the prospect of stronger statements.

In our derivation of the likelihood we assume that the noise in each
detector is Gaussian and stationary.  While the fundamental noise
sources in gravitational wave detectors are all characterized by
Gaussian-stationary statistics, the realities of an actual
implementation --- e.g., detector imperfections and environmental
couplings --- guarantee that the actual noise character will be
neither Gaussian nor stationary.  Characterizing the detector, which
includes identifying instrumental and environmental artifacts in the
``gravity-wave channel'' and regressing or removing these artifacts
to the greatest possible extent, is a necessary pre-requisite to the
analysis of the data for gravitational waves.  We assume here that
these artifacts have already been dealt with as best possible, so that
$\bbox{g}$ contains no identifiable instrumental artifacts, transient
or otherwise, and that the noise is stationary on timescales long
compared to the duration of a signal.

Even so, the noise will remain non-Gaussian and non-stationary.  As
long as the evolution of the noise character is adiabatic --- i.e., it
varies only on timescales long compared to the signal duration and the
time required to estimate the moments of the noise distribution --- we
can treat the noise as stationary on any important sub-interval.  It
is also an excellent approximation to treat the noise as Gaussian.  To
see why, focus on a stationary noise process but without any
assumption on its distribution ${\cal P}(x)$.  Suppose, by
observations on the noise, we estimate its mean $\mu$ ($\overline{x}$)
and variance $\sigma^{2}$ ($\overline{(x-\mu)^{2}}$).  What
probability distribution makes the least assumptions about the process
while remaining consistent with this information?  This
information-theoretic question has a definite answer: it is the
probability distribution $P(x|I)$ with maximum entropy, subject to the
constraints on its mean and variance: i.e., the distribution
$P(x|\mu,\sigma)$ that maximizes the functional
\begin{eqnarray}
\lefteqn{I(P) := -\int_{-\infty}^\infty dx\;P(x|\mu,\sigma)\ln 
P(x|\mu,\sigma)
-\lambda_0 \int_{-\infty}^\infty dx\; P(x|\mu,\sigma)}&&\nonumber\\
&&-\lambda_1 \int_{-\infty}^\infty dx\;x\, P(x|\mu,\sigma)
-\lambda_2 \int_{-\infty}^\infty dx\;(x-\mu)^2\, P(x|\mu,\sigma)
\end{eqnarray}
where $\lambda_0$, $\lambda_1$ and $\lambda_2$ are the Lagrange
multipliers corresponding to the constraints on the overall
normalization of the probability, the mean and the variance.  The
solution to this variational problem is
\begin{equation}
P(x|I) =
{\exp\left[(x-\mu)^2/2\sigma^2\right]\over\sqrt{2\pi\sigma^2}},
\end{equation}
i.e., the normal distribution.  Similarly, if we have a
correlated process known to us only through its mean and
autocorrelation,
\begin{equation}
    C[k-j] := \overline{\left(x[j]-\mu\right)\left(x[k]-\mu\right)},
\end{equation}
the distribution that makes the least assumptions about the process
beyond these is the multivariate normal distribution
\begin{equation}
    P\left(\left\{x[j]\right\}|\mu,C\right) = 
    {
    \exp\left[
    -{1\over2}\sum_{j,k=0}^{N_{T}-1}\left(x[k]-\mu\right)
    \left|\left|\bbox{C}^{-1}\right|\right|_{kl}
    \left(x[l]-\mu\right)
    \right]
    \over
    \sqrt{\left(2\pi\right)^{N_{T}}\text{det}||\bbox{C}||}
    },
\end{equation}
where $\left\{x[k]\right\}$ is a set of $N_{T}$ samples indexed by the
sample times and the matrix $\bbox{C}$ has components
\begin{equation}
    \left|\left|\bbox{C}\right|\right|_{jk} := C[k-j]. 
\end{equation}
Thus, modeling (correlated) noise as arising from a (multi-variate)
Normal distribution is simultaneously the best and most conservative
assumption that one can make when all one knows is the noise mean and
(co-)variance.

Put another way, if our only knowledge of the noise character is its
first and second moments (i.e., mean and correlation function or
power spectrum) then treating the noise as anything but Normally
distributed is to assume \textit{more} than we actually know and,
consequently, would lead us to \textit{overstate} the accuracy of any
conclusions we reach.

We choose a particular normalization of the likelihood function,
as the ratio of $P(\bbox{g}|\bbox{\theta})$ to the probability that
$\bbox{g}$ arises from detector noise alone.  The signal is assumed to
be a plane gravitational wave incident on the detector array, so that
the detectors respond to the signal coherently.  If the hypothesized
signal is characterized by parameters $\bbox{\theta}$ (e.g.,
binary system chirp mass, orientation, distance, etc.) then we
denote the likelihood function $\Lambda(\bbox{g}|\bbox{\theta})$ and
regard it, for fixed receiver output $\bbox{g}$, as a function of the
signal parameterization $\bbox{\theta}$.

The derivation of $\Lambda(\bbox{g}|{\bbox{\theta}})$ for a
deterministic signal in a multi-detector receiver parallels closely
the derivation of $\Lambda(g|{\bbox{\theta}})$ given in
\cite[\S{II}]{finn92a} for a single-detector receiver; however, in the
generalization several new elements arise and are discussed here for
the first time. 

In section \ref{sec:p(g|0)} we walk through the construction of
$P(\bbox{g}|H_{0},{\cal I})$, the probability of observing $\bbox{g}$
in the absence of any signal.  Evaluation of
$P(\bbox{g}|H_{\bbox{\theta}},{\cal I})$ (the probability of observing
$\bbox{g}$ when the signal characterized by $\bbox{\theta}$ is
present) and the likelihood function itself proceed much more quickly
in section \ref{sec:LambdaDet}.  In section \ref{sec:mlt} we describe
a detection test based solely on the maximum of the likelihood and
identify the signal-to-noise ratio with the maximum of $\ln\Lambda$
over $\bbox{\theta}$.  In section \ref{sec:fast<a,b>} we discuss how
the likelihood function can be evaluated quickly and efficiently,
which is critical to its use in real data analysis.

\subsection{Probability that $\bbox{g}$ is receiver
noise}\label{sec:p(g|0)} 

Central to the evaluation of $\Lambda(\bbox{g}|\bbox{\theta})$ is the
evaluation of $P(\bbox{g}|H_0,{\cal I})$, the probability that the
receiver output $\bbox{g}$ is an example of receiver noise $\bbox{n}$.
(Here $\cal I$ denotes other, unenumerated conditions.) The
formulation of this probability density for a single detector
(scalar-valued time series $g$) was discussed in \cite{finn92a}; in
this section we review that discussion, setting the stage for our
treatment of the more general problem of a multi-channel receiver
(vector-valued $\bbox{g}$).

\subsubsection{Single-channel time series}

Focus attention first on the single-channel output of a single
detector.  When the detector noise is Gaussian and stationary, any
single sample $n[j]$ of detector noise is normally distributed with a
mean and variance independent of when the sample was taken. 
Without loss of generality we can assume that the ensemble mean
$\overline{{n}}$ vanishes, in which case the probability that a sample
$g[j]$ of detector output is a sample of detector noise is
\begin{mathletters}
\begin{equation}
P(g[j]|H_0,{\cal I}) = {
\exp\left[-g[j]^2/2\sigma^2\right]
\over\sqrt{2\pi\sigma^2}
} . \label{eq:p(g[j]|0)}
\end{equation}
The variance $\sigma^2$ of the distribution is the ensemble 
average of the square of the detector noise:
\begin{equation}
\sigma^2 = \overline{n^2}.
\end{equation}
\end{mathletters}

Equation \ref{eq:p(g[j]|0)} holds true for each sample $g[j]$;
consequently, the joint probability that the length $N_T$ 
sequence $g$ is a sample of detector noise is given by the 
multivariate Gaussian distribution
\begin{equation}
P(g|H_0,{\cal I}) =
{\exp\left[-{1\over2}\sum_{j,k=0}^{N_T-1}
g[j]\left|\left|\bbox{C}^{-1}\right|\right|_{jk}g[k]\right]\over
\sqrt{\left(2\pi\right)^{N_T}\det\left|\left|\bbox{C}\right|\right|}}
\label{eq:p(g|0)}
\end{equation}
In place of the variance $\sigma^2$ that appears in the exponent of
equation \ref{eq:p(g[j]|0)} is the matrix $\bbox{C}^{{-1}}$.  As
$P(g|H_{0},{\cal I})$ is a probability so $\bbox{C}^{{-1}}$ is
positive definite and invertible. The matrix $\bbox{C}$ gives the 
covariance of the noise process: it is equal to the ensemble average 
of the product of the detector noise at different samples:
\begin{equation}
\overline{{n}[j]{n}[k]} = \left|\left|\bbox{C}\right|\right|_{jk}.
\end{equation}
Since the detector noise is also assumed to be stationary,
$\left|\left|\bbox{C}\right|\right|_{jk}$ can depend only on the
difference $j-k$; correspondingly, $\bbox{C}$ is constant on its
diagonals: i.e., it is a \textit{Toeplitz} matrix.  Since
$\bbox{C}$ is Toeplitz it is characterized completely by the scalar
sequence $c[k]$ of length $2N_T-1$ whose elements are the first row
and column of $\bbox{C}$:
\begin{equation}
c[j-k] := \left|\left|\bbox{C}\right|\right|_{jk}. 
\end{equation}
This sequence is the noise autocorrelation. The DFT of $c[k]$ is 
related to the noise power spectral density. 

\subsubsection{Multi-channel time series}
Now consider a receiver consisting of $N_D$ component detectors, where
the receiver output is an $N_D$-dimensional time series.  Without loss
of generality assume that, while the different detectors that comprise
the receiver may be sampled at different rates, all the detector
outputs have been resampled so that the interval between samples in
all detector data streams is $\Delta t$.  The receiver output
$\bbox{g}$ is a multi-channel time series: a vector-valued sequence
consisting of the direct sum of the output $g_{\alpha}$ of the
individual detectors.  Similarly, the receiver noise $\bbox{n}$ is the
direct sum of the noise contributions $n_\alpha$ to the output of the
$N_D$ detectors:
\begin{mathletters}
\begin{eqnarray}
\bbox{g}[k] &:=& g_1[k]\oplus \cdots \oplus g_{N_D}[k]\\
\bbox{n}[k] &:=& n_{1}[k]\oplus \cdots \oplus n_{N_D}[k].
\end{eqnarray}
\end{mathletters}

Focus attention on a single sample of the receiver output: i.e.,
the vector $\bbox{g}[j]$.  Since the receiver noise is Gaussian and
stationary, the probability that sample $\bbox{g}[j]$ is a sample of
receiver noise $\bbox{n}$ is given by the multivariate Gaussian
\begin{mathletters}
\label{eq:p(bg[j]|0)}
\begin{equation}
P(\bbox{g}[j]|H_0,{\cal I}) =
{\exp\left[
-{1\over2}\bbox{g}[j]\cdot\bbox{C}[0]^{-1}\cdot\bbox{g}[j]
\right]\over\sqrt{\left(2\pi\right)^{N_D}
\det\left|\left|\bbox{C}[0]\right|\right|}}
\end{equation}
where the matrix $\bbox{C}[0]$ is the ensemble average of the outer
product of a sample of receiver noise with itself:
\begin{equation}
\bbox{C}[0]  = \overline{\bbox{n}[j]\otimes\bbox{n}[j]}. 
\end{equation}
\end{mathletters}

As suggested by the selection index $[0]$, $\bbox{C}[0]$ is one
element of a sequence of correlation matrices.  Each element of this
sequences arises from the ensemble average of the outer product of
receiver noise with itself at a different time.  Since the receiver
noise is stationary, the ensemble average depends only on the time
difference; consequently,
\begin{equation}
\bbox{C}[j-k] = \overline{\bbox{n}[j]\otimes\bbox{n}[k]}.
\label{eq:Cn}
\end{equation}
Just as the DFT of the autocorrelation sequence for a single detector 
is related to its noise power spectral density, so the DFT of 
$\bbox{C}[k]$ is related to its noise power spectral density. In this 
case, the power spectral density is a matrix-valued function of 
frequency, with the diagonal components equal to the power spectral 
density of the noise in a particular detector and the off-diagonal 
components related to the cross-spectral density of the noise in two 
different detectors. 

Equations \ref{eq:p(bg[j]|0)} hold true for each individual sample of 
receiver output $\bbox{g}[j]$; consequently, the joint probability 
that the length $N_T$ sequence $\bbox{g}$ is a sample of receiver 
noise is also a multivariate Gaussian:
\begin{mathletters}
\begin{equation}
P(\bbox{g}|H_0,{\cal I}) 
= {\exp\left[
-{1\over2}
\sum_{j,k=0}^{N_T-1}
\bbox{g}[j]\cdot
\left|\left|{\Bbb T}_{N_T}(\bbox{C})^{-1}\right|\right|_{jk}\cdot
\bbox{g}[k]\right]\over
\sqrt{\left(2\pi\right)^{N_TN_D}D_{N_T}(\bbox{C})}}
\label{eq:p(bg|0)},
\end{equation}
where 
${\Bbb T}_{N_T}(\bbox{C})$ is a \textit{block Toeplitz matrix,} i.e., 
a matrix whose elements
$\left|\left|{\Bbb T}_{N_T}(\bbox{C})\right|\right|_{jk}$ 
are themselves Toeplitz matrices:
\begin{equation}
\left|\left|{\Bbb T}_{N_{T}}(\bbox{C})\right|\right|_{jk} = 
\bbox{C}[j-k].
\end{equation}
Each ``element'' of ${\Bbb T}_{N_{T}}(\bbox{C})$ is thus a
$N_{D}\times N_{D}$ matrix.  The inverse ${\Bbb
T}_{N_{T}}(\bbox{C})^{-1}$ of ${\Bbb T}_{N_{T}}(\bbox{C})$
is defined in the usual way
\begin{eqnarray}
\delta_{jk}{\Bbb I}_{N_{D}} &=& \sum_{l=0}^{N_{T}-1}
\left|\left|{\Bbb T}_{N_{T}}(\bbox{C})\right|\right|_{jl}\cdot
\left|\left|{\Bbb T}_{N_{T}}(\bbox{C})^{-1}\right|\right|_{lk}\\
&=& \sum_{l=0}^{N_{T}-1} \bbox{C}[j-l]\cdot
\left|\left|{\Bbb T}_{N_T}(\bbox{C})^{-1}\right|\right|_{lk},
\end{eqnarray}
where $\cdot$ indicates the usual matrix product between $N_{D}\times 
N_{D}$ matrices and ${\Bbb I}_{N_{D}}$ is the $N_{D}\times N_{D}$ 
unity matrix.

Finally, 
\begin{equation}
D_{N_{T}}(\bbox{C}) = \det {\Bbb T}_{N_{T}}(\bbox{C});
\end{equation}
\end{mathletters}
i.e, $D_{N_{T}}(\bbox{C})$ is the determinant of ${\Bbb 
T}_{N_{T}}(\bbox{C})$, regarded as an $N_{T}N_{D}\times N_{T}N_{D}$ 
matrix. 

The argument of the exponential in equation \ref{eq:p(bg|0)} takes the
general form of a symmetric inner product of two vector-valued sequences
with respect to ${\Bbb T}_{N_T}(\bbox{C})^{-1}$. Terms like this occur
frequently enough that it is convenient to introduce a special notation for
them: in particular, we define the \textit{symmetric} inner product
\begin{equation}
<\bbox{a},\bbox{b}>_{\bbox{C}} := 
{1\over4}
\sum_{j,k=0}^{N_{T}-1}
\left[
\bbox{a}[j]\cdot
\left|\left|{\Bbb T}_{N_T}(\bbox{C})^{-1}\right|\right|_{jk}
\cdot\bbox{b}[k]+
\bbox{b}[j]\cdot
\left|\left|{\Bbb T}_{N_T}(\bbox{C})^{-1}\right|\right|_{jk}
\cdot\bbox{a}[k]
\right]
\label{eq:<>}
\end{equation}
so that
\begin{equation}
P(\bbox{g}|0,{\cal I}) = {
\exp\left[-<\bbox{g},\bbox{g}>_{\bbox{C}}\right]\over
\sqrt{\left(2\pi\right)^{N_TN_D}D_{N_T}(\bbox{C})}}.
\end{equation}

\subsection{The likelihood function for a deterministic
signal}\label{sec:LambdaDet} 

Consider the case of a \textit{deterministic signal,} i.e., a
signal whose time series is known up to parameters $\bbox{\theta}$.
An example is the gravitational-wave signal from a coalescing binary
system, where the parameters include the binary's position and
orientation on the sky, component masses and spins, and distance.

Distinguish between the signal itself, which we denote
$\bbox{h}({\bbox{\theta}})$, and the receiver response to the signal,
which we denote $\bbox{m}({\bbox{\theta}})$.  Assume that the receiver
response is linear in the applied signal; consequently, the
probability of observing $\bbox{g}$ when the signal
$\bbox{h}({\bbox{\theta}})$ is present on the receiver is the same as
the probability of observing $\bbox{g}-\bbox{m}({\bbox{\theta}})$ as a
sample of receiver noise:
\begin{mathletters}
\label{eq:p(g|theta,gauss,det)}
\begin{eqnarray}
P(\bbox{g}|H_{\bbox{\theta}},{\cal I}) &=&
P\left(\bbox{g}-\bbox{m}({\bbox{\theta}})|H_0,{\cal I}\right)\\
&=&
{
\exp\left[
-\left<\bbox{g}-\bbox{m}({\bbox{\theta}}),
\bbox{g}-\bbox{m}({\bbox{\theta}})\right>_{\bbox{C}}
\right]\over
\sqrt{\left(2\pi\right)^{N_DN_T}D_{N_T}(\bbox{C})}
}
\end{eqnarray}
\end{mathletters}
The likelihood function is thus
\begin{eqnarray}
\Lambda(\bbox{g}|\bbox{\theta}) &=&
\exp\left[
2\left<\bbox{g},\bbox{m}(\bbox{\theta})\right>_{\bbox{C}}
-\left<\bbox{m}(\bbox{\theta}),\bbox{m}(\bbox{\theta})
\right>_{\bbox{C}}\right],
\label{eq:LambdaDetSig}
\end{eqnarray}
where we have exploited the symmetry of the inner product $<>$ (cf.\
eq.~\ref{eq:<>}). 

The sole influence of the signal on the likelihood owes to the first
term in the argument of the exponential in equation
\ref{eq:LambdaDetSig},
$\left<\bbox{g},\bbox{m}(\bbox{\theta})\right>_{\bbox{C}}$.  This term
is a linear filter applied to the observation $\bbox{g}$.  That linear
filter is the Wiener optimal filter for a gravitational wave receiver.

\subsection{The maximum likelihood test}\label{sec:mlt}

In a Bayesian analysis, the product of the likelihood and the \textit{a
priori} probability density $P(\bbox{\theta}|{\cal I})$ is
proportional to the \textit{a posteriori} probability
$P(\bbox{\theta}|\bbox{g})$.  From this probability density one can
decide with what confidence one believes that a signal has been
observed or construct Bayesian credible sets --- regions of parameter
space encompassing a given fraction of the total probability that
$\bbox{\theta}$ takes on a given value. 

In a frequentist analysis confidence intervals and upper limits are
constructed from the likelihood, although certain \textit{ad hoc}
assumptions are required to make the procedure definite
\cite{feldman98a,giunti98b,giunti98a,roe99a}.  When our goal is simply
to decide whether a signal is present a common candidate procedure,
recommended both by analysis \cite{poor94a} and intuitive
appeal, is the maximum likelihood test: Let
\begin{mathletters}
\begin{eqnarray}
\bbox{\Theta} &=& 
\left(\begin{array}{l}
\text{Domain of physically permissible $\bbox{\theta}$}
\end{array}\right),\\
\Lambda_{\max}(\bbox{g}) &=&
\max_{\bbox{\theta}\in\bbox{\Theta}}\Lambda(\bbox{g}|\bbox{\theta}),\\
\widehat{\bbox{\theta}}(\bbox{g}) &=& 
\left(\begin{array}{l}
\text{Parameter $\bbox{\theta}$ for which}\\
\text{$\Lambda(\bbox{g}|\bbox{\theta}) = \Lambda_{\max}(\bbox{g}).$} 
\end{array}\right),
\end{eqnarray}
\end{mathletters}
and choose a threshold $\Lambda_0$. Given an observation $\bbox{g}$,
evaluate $\Lambda_{\max}(\bbox{g})$. If $\Lambda_{\max}$ exceeds
$\Lambda_0$, then conclude that the signal corresponding to
$\widehat{\bbox{\theta}}(\bbox{g})$ has been observed.

If $\widehat{\bbox{\theta}}$ is not on the boundary of $\bbox{\Theta}$,
then the maximum value of the likelihood function is also an extremum
of $\Lambda(\bbox{g}|\bbox{\theta})$ and
\begin{equation}
\ln\Lambda(\bbox{g}|\widehat{\bbox{\theta}}) =
<\bbox{m}_{\widehat{\bbox{\theta}}}, \bbox{m}_{\widehat{\bbox{\theta}}}>.
\end{equation}

\subsection{Evaluation of
  $<\bbox{a},\bbox{b}>_{\bbox{C}}$}\label{sec:fast<a,b>}  

A naive evaluation of $<\bbox{a},\bbox{b}>_{\bbox{C}}$ following its
definition in equation \ref{eq:<>} has a high computational cost: 
\begin{enumerate}
\item Solving the linear system ${\Bbb T}_{N_T}(\bbox{C})\cdot\bbox{x} =
\bbox{b}$ requires ${\cal O}[(N_TN_D)^3]$ operations;
\item Evaluating the inner product $\bbox{a}\cdot\bbox{x}$
requires ${\cal O}(N_TN_D)$ operations.
\end{enumerate}
The operation count for this evaluation of
$<\bbox{a},\bbox{b}>_{\bbox{C}}$ is dominated by the solution of the
linear system in the first step and would appear to be prohibitive,
even if if could be done accurately, for all but the shortest time
series.

In fact $<\bbox{a},\bbox{b}>_{\bbox{C}}$ can be evaluated in at most
${\cal O}(N_{D}^{2}N_T)$ operations.
To do so requires only that we
pre-process the input data through a chain of between one and three
linear filters.  In this section we describe these filters and the
inner-product of the filtered time-series.

\subsubsection{The Linear Filter}\label{sec:preprocess}

The desired pre-processing is conveniently described as a sequence of
three linear filters.  The first filter simply whitens separately the
output of each detector in the receiver.  The second forms linear
combinations of the whitened detector outputs to form a basis of
``pseudo-detectors'' whose cross-correlations vanishes, while the
third whitens the pseudo-detector output.  Thus, the first filter can
be formed and applied without reference to the other detectors, while
the second and third filters are the identity if the cross-correlation
between distinct detectors vanishes.

\paragraph{First filter: Whiten detector noise} Using any convenient
technique \cite{ljung99a} whiten the output of each detector.  
Denote with a prime the whitened detector outputs and their
cross-correlation when regarded as part of a receiver: i.e.,
$n'_{k}$ is the whitened noise from detector $k$ and
\begin{equation}
    \bbox{C}^{\prime}[j-k] = \overline{\bbox{n}'[j]\otimes\bbox{n}'[k]}. 
\end{equation}

Write the cross-spectral density Toeplitz matrix $\bbox{C}'$ in
block-form as
\begin{mathletters}
\begin{equation}
    \bbox{C}' = \left(
    \begin{array}{cccc}
        \bbox{C}^{\prime}_{11}&\bbox{C}^{\prime}_{12}&\dots&
        \bbox{C}^{\prime}_{1N_{D}}\\
        \bbox{C}^{\prime}_{21}&\bbox{C}^{\prime}_{22}&\dots&
        \bbox{C}^{\prime}_{2N_{D}}\\
        \vdots&\vdots&\ddots&\vdots\\
        \bbox{C}^{\prime}_{N_{D}1}&\bbox{C}^{\prime}_{N_{D}2}&\dots&
        \bbox{C}^{\prime}_{N_{D}N_{D}}
    \end{array}
    \right),\label{eq:c'}
\end{equation}
where
\begin{equation}
    C^{\prime}_{jk}[l-m] = \overline{n^{\prime}_{j}[l]n^{\prime}_{k}[m]}
\end{equation}
\end{mathletters}
Focus on the diagonal blocks $\bbox{C}^{\prime}_{kk}$, which are the
Toeplitz matrices corresponding to the autocorrelation of the output
of detector $k$.  These blocks are constant multiples of the unity
matrix, the constant being simply the mean-square noise amplitude in
detector $k$, $\sigma^{2}_{k}$.  Absorb this constant into the
whitening filter for each detector so that the whitened output has
mean-square amplitude unity and the $\bbox{C}^{\prime}_{{kk}}$ are 
just the unity matrix. 

Focus attention now on cross-correlations, which are represented by
the off-diagonal Toeplitz matrices in equation \ref{eq:c'}.  If the
cross-correlation between the detector outputs is consistent with zero
then we are done with the pre-processing.  If, on the other hand,
$\bbox{C}'[j-k]$ is non-zero for $j\neq k$, then we have two
additional pre-processing steps, which we describe below.

\paragraph{Second filter: Diagonalization} The vector 
$\bbox{g}^{\prime}$ corresponds to the direct sum of the output of the 
several detectors that form the network, after their output has been 
separately whitened. In this step we 
we form a new ``basis'' 
of detectors whose noise is uncorrelated: i.e., we find the 
linear filter described by the coefficients $\bbox{b}[k]$ and 
$\bbox{a}[k]$ such that $\bbox{n}^{\prime\prime}$, 
\begin{equation}
    \bbox{n}^{\prime\prime}[k] := 
    \sum_{j=0}^{N_{b}}\bbox{b}[{j}]\cdot\bbox{n}^{\prime}[k-j]
    - \sum_{j=1}^{N_{a}}\bbox{a}[{j}]\cdot\bbox{n}^{\prime\prime}[k-j],
\end{equation}
has the property
\begin{equation}
    \overline{
    {n}^{\prime\prime}_{j}[l]
    \otimes
    {n}^{\prime\prime}_{k}[m]
    } = C^{\prime\prime}_{j}[l-m]\delta_{jk}.
\end{equation}
This transformation, applied to $\bbox{g}^{\prime}$, yields
$\bbox{g}^{\prime\prime}$.

Standard system identification techniques \cite{ljung99a} can be used
to find the appropriate transformation. For purposes of illustration
only we describe one way to find such a transformation, involving just
sequence $\bbox{b}$. Focus on the discrete Fourier transform of
$\bbox{C}'[k]$,
\begin{equation}
    \widetilde{\bbox{C}^{\prime}}[k] = 
    \sum_{n = -N}^{N} \bbox{C}^{\prime}[j]
    \omega_{2N-1}^{jk}, 
\end{equation}
where we have chosen $N$ large enough that $\bbox{C}^{\prime}[k]$
vanishes for $k>N$.  The quantity $\widetilde{\bbox{C}^{\prime}}$ is
the two-sided (discrete) cross-spectral density of $\bbox{n}'$.  Each
component $\widetilde{\bbox{C}^{\prime}}[k]$ is an $N_{D}\times N_{D}$
Hermitian matrix.  Consider the sequence of unitary transformations
$\widetilde{\bbox{U}}[k]$ such that
\begin{equation}
    \widetilde{\bbox{U}}[k]^{\dagger}
    \cdot\widetilde{\bbox{C}^{\prime}}[k]
    \cdot\widetilde{\bbox{U}}[k]
\end{equation}
are each diagonal.  This sequence exists as long as the noise is not
fully correlated in any pair of detectors, at any frequency. 
Additionally, the symmetries of $\bbox{C}$ guarantee that
$\widetilde{\bbox{U}}[k]$ equals $\widetilde{\bbox{U}}^{\dagger}[-k]$. 
Consequently, the inverse discrete Fourier transform of the sequence
$\widetilde{\bbox{U}}[k]$ yields a real linear filter $\bbox{U}[k]$
that, when applied to the vector-valued receiver noise
$\bbox{n}'$, yields an output $\bbox{n}''$ whose cross-correlation is
diagonal: i.e.,
\begin{mathletters}
    \begin{equation}
\overline{n^{\prime\prime}_{j}[l]n^{\prime\prime}_{k}[m]} = 
0\qquad\text{if $j\neq k$}
\end{equation}
where
\begin{equation}
    \bbox{n}''[k] = \sum_{j=-N}^{N}\bbox{U}[j]\cdot\bbox{n}'[k-j].
\end{equation}
\end{mathletters}

\paragraph{Third filter: Final whitening} Following the formation of a
pseudo-detector basis the $n^{\prime\prime}_{j}$ will not necessarily
be white.  The final step of pre-processing is to whiten separately
the output sequence corresponding to each pseudo-detector, absorbing
the overall normalization $\sigma^{\prime\prime\prime}_{j}$ into the
filter so that the rms output of each pseudo-detector is unity.  This
final step, since it does not involve combining the output of the
different pseudo-detectors, does not change the vanishing
cross-correlation of the output of different pseudo-detectors.

Following this final pre-processing we are left with the receiver
output $\bbox{g}'''$ whose noise component has the desired property
\begin{equation}
    \overline{
    \bbox{n}^{\prime\prime\prime}_{j}[l]
    \otimes\bbox{n}^{\prime\prime\prime}_{k}[m]
    } = \delta_{jk}\delta_{lm};
\end{equation}
i.e., the noise in each pseudo-detector is white, and the noise in 
different pseudo-detectors is uncorrelated. 

Since all of the operations involved in this pre-processing are linear
filter operations the computational cost of processing a length
$N_{T}$ sequence of network output $\bbox{g}$ to $\bbox{g}''$ is
strictly proportional to $N_{T}$.  The order of the filters involved
is \textit{independent} of $N_{T}$.  The determination of the filters
themselves may be a somewhat time-consuming operation; however, since
the receiver noise is stationary on timescales long compared to the
signal duration these filters need to be found or modified very
infrequently.\footnote{Were this not true then we would fail also to
satisfy the basic assumptions that allow us to construct the optimal
filter for a single detector.} The first and (if needed) third linear
filter operations are done separately for each detector or
pseudo-detector; correspondingly, they involve a factor of $N_{D}$. 
Finally, the second transformation, which (if needed) forms the basis
of independent detectors, is really ${\cal O}(N_{D}^{2})$ linear
filters.  Hence, evaluation of the linear filter on a sequence of
length $N_{T}$ requires ${\cal O}(N_{D}^{2}N_{T})$ operations
generally, falling to ${\cal O}(N_{D}N_{T})$ if there is no
cross-correlated noise between the detectors.

\subsubsection{Reformulating the inner product}

We presume that we have pre-processed the receiver output as described
in the sec.~\ref{sec:preprocess}. Following this preprocessing the
inner product can be rewritten as
\begin{eqnarray}
<\bbox{a},\bbox{b}>_{\bbox{C}} &:=&
{1\over4}
\sum_{j,k=0}^{N_{T}-1}
\left[
\bbox{a}[j]\cdot
\left|\left|{\Bbb T}_{N_T}(\bbox{C})^{-1}\right|\right|_{jk}
\cdot\bbox{b}[k]+
\bbox{b}[j]\cdot
\left|\left|{\Bbb T}_{N_T}(\bbox{C})^{-1}\right|\right|_{jk}
\cdot\bbox{a}[k]
\right]\\
&=&
{1\over2}
\sum_{j,k=0}^{N_{T}-1}
\bbox{a}''[j]\cdot\bbox{b}''[j], 
\end{eqnarray}
which has an operation count of ${\cal O}(N_{D}N_{T})$.  Including the
pre-processing, the operation count scales linear with $N_{T}$ and
either linearly or, at most, quadratically with the number of
detectors in the receiver $N_{D}$.

Most of the work involved in calculating the inner product is in
dealing with the correlations: both auto-correlations of the
individual detector outputs and cross-correlations of the different
detectors in the receiver: i.e., in ``inverting'' ${\Bbb
T}_{N_T}(\bbox{C})$.  Owing to the special structure of ${\Bbb
T}_{N_T}(\bbox{C})$ the effect of its inverse in the inner product can
be expressed by applying a sequence of linear filters, each of order
independent of $N_{T}$, to the inputs (i.e., the linear
transformations described in sec.~\ref{sec:preprocess}).  These
transformations are determined entirely by the statistical character
of the noise in the receiver, which changes only on timescales long
compared to the signal duration; consequently, the asymptotic
operation count for the inner product is at most ${\cal
O}(N_{T}N_{D}^{2})$ and not ${\cal O}[(N_{T}N_{D})^{3}]$ as a naive
estimate might suggest.

\subsection{Signal-to-noise}\label{sec:snr}

Following the identification of pseudo-detectors whose noise is white
and uncorrelated the inner product
$<\bbox{m}_{\widehat{\bbox{\theta}}},
\bbox{m}_{\widehat{\bbox{\theta}}}>$, which is the maximum of the
log-likelihood, is recognized as the half the sum, over the
pseudo-detectors, of the ratio of two quantities: the mean-square
response of the receiver to the signal and the mean-square
pseudo-detector noise.  Correspondingly, we identify
\begin{equation}
\rho^2 := <\bbox{m}_{\widehat{\bbox{\theta}}},
\bbox{m}_{\widehat{\bbox{\theta}}}>
\end{equation}
as the (power, or amplitude-squared) signal-to-noise ratio.

When the detector noise is uncorrelated the signal-to-noise of the
network is clearly related to the sum of the signal-to-noise of the
component detectors: related --- not equal --- because in the analysis
of several detectors as a single receiver we have imposed the
important constraint that the signal parameters appearing in the
separate signal-to-noise ratios are identical.  To see the importance
of this constraint, consider $N_D$ detectors making an observation for
a signal $\bbox{s}(t|\theta)$.  Assume for simplicity that everything
about the signal (including its waveform and start time) is known
except for its amplitude $\theta$, so that
\begin{equation}
    \bbox{s}(t|\theta) := \theta\bbox{s}(t).
\end{equation}
Suppose also that the noise in each detector is uncorrelated with the
noise in any other detector.  Since the detectors are independent in
this way we can write the receiver likelihood
$\Lambda(\bbox{g}|\theta)$ as a product of the separate detector
likelihoods $\Lambda_j(g_j|\theta)$.

Now consider detector $j$ acting independently of the rest.  For this
detector,
\begin{equation}
  \ln\Lambda_j(g_j|\theta) = 
2\theta\left<g_j,s_j\right>_{C_j} - 
\theta^2\left<s_j,s_j\right>_{C_j},
\end{equation}
where the subscript $j$ indicates the relevant quantity with regards
to detector $j$. The maximum likelihood point-estimate of $\theta$
based on the observation $g_j$ in detector $j$ is
\begin{equation}
  \theta_j = {\left<g_j,s_j\right>_{C_j}
    \over\left<s_j,s_j\right>_{C_j}}
\end{equation} 
and the corresponding S/N is
\begin{mathletters}
\begin{eqnarray}
  \rho_j^2 &=& 2\theta_j^2\left<s_j,s_j\right>_{C_j}\\
  &=& 2{\left<g_j,s_j\right>^2_{C_j}
    \over\left<s_j,s_j\right>_{C_j}}.
\end{eqnarray}
\end{mathletters}
In general, $\theta_j$ will not be equal to $\theta_k$ and $\rho_j$
will not be equal to $\rho_k$.  If $g_{j}$ is (Gaussian-stationary)
receiver noise, than the ensemble average of each $\rho^{2}_{j}$ is
equal to unity.  The maximum of the log of the product of the
likelihoods for the separate detectors is then half the sum over the
$\rho^{2}_{j}$, or $N_{D}/2$.

On the other hand, if we consider all the detectors to be part of a
network, then
\begin{equation}
  \ln\Lambda(\bbox{g}|\theta) = 
  2\theta\left<\bbox{g},\bbox{s}\right>_{\bbox{C}}-
  \theta^2\left<\bbox{s},\bbox{s}\right>_{\bbox{C}}.
\end{equation}
The maximum likelihood point-estimate of $\theta$ based on the
observation $\bbox{g}$ is 
\begin{equation}
  \theta = {\left<\bbox{g},\bbox{s}\right>_{\bbox{C}}
    \over\left<\bbox{s},\bbox{s}\right>_{\bbox{C}}}
\end{equation} 
and the corresponding S/N is
\begin{mathletters}
\begin{eqnarray}
  \rho^2 &=& 2\theta^2\left<\bbox{s},\bbox{s}\right>_{\bbox{C}}\\
  &=& 2{\left<\bbox{g},\bbox{s}\right>^2_{\bbox{C}}
    \over\left<\bbox{s},\bbox{s}\right>_{\bbox{C}}}\\
  &=&
  2{\sum_j\left<g_j,s_j\right>^2_{{C}_j}
    \over\left<\bbox{s},\bbox{s}\right>_{\bbox{C}}}
\end{eqnarray}
\end{mathletters}
The S/N $\rho^2$ for the network is not equal to the S/N in any
component detector, nor is it equal to the sum of the $\rho_j^2$
treated as independent quantities.  With the constraint that the
individual detectors in the receiver must respond coherently to any
signal, the ensemble average of $\rho^{2}$ is equal to unity, not
$N_{D}$.  The constraint that each detector in the receiver responds
coherently to the incident signal reduces the variance in $\rho^2$
below what one would expect from a simple sum of the individual
$\rho_j^2$.

\section{Example applications}
\label{sec:examples}

\subsection{Introduction}

In this section we apply the formalism developed in the previous two
sections in several numerical examples, which are based on a model
source detected by a model receiver.  Our aim is to illustrate one way
that likelihood-based detection might be used in a network analysis,
to demonstrate that its performance is superior to the kind of
coincidence based analysis described by
\cite{jaranowski94a,jaranowski96a,creighton99a}, and to explore the
performance of the likelihood and coincidence tests when the detector
noise is strongly non-Gaussian.

The model source and detectors are described in section
\ref{sec:toy}. For simplicity we focus on testing the null hypothesis
$H_0$ (i.e., ``the signal is absent'').  In section
\ref{sec:FreqTests} we describe two different ways of testing this
hypothesis: via a threshold placed on the maximum of the likelihood
function for the joint output of all the detectors and via an analysis
of ``coincidences'' between events identified separately in each
detector. Monte Carlo simulations are used to evaluate the detection
efficiency as a function of the false alarm fraction for each test in
two different circumstances: Gaussian detector noise and a
mixture-Gaussian model of non-Gaussian detector noise.

\subsection{Model receiver and source model}\label{sec:toy}

Consider two detectors, denoted ``${}+{}$'' and ``${}-{}$'', separated 
by a distance $2R$.  For the purpose of illustration, assume that the 
noise in each detector is white (i.e., uncorrelated) up to the 
Nyquist frequency, uncorrelated between the two detectors, and has 
two-sided power spectral density amplitudes $S_{+}$ and $S_{-}$ in the 
${}+{}$ and ${}-{}$ detectors, respectively.  Assume also that these 
detectors have no orientation: they respond identically to radiation 
incident from any direction. 

In addition to parameters that describe the internal state of a 
radiation source and its orientation relative to the detector line of 
sight, every signal incident on the receiver is characterized by a 
``signal start time'', describing when the initial wavefront reaches
the receiver.  It is convenient to measure time at the mid-point between 
the two detectors in our receiver, so that the signal start time is 
defined to be the moment that the initial wavefront reaches the 
mid-point between the two detectors. 

\begin{figure}
  \epsfxsize=0.75\columnwidth
  \begin{center}
    \leavevmode\epsffile{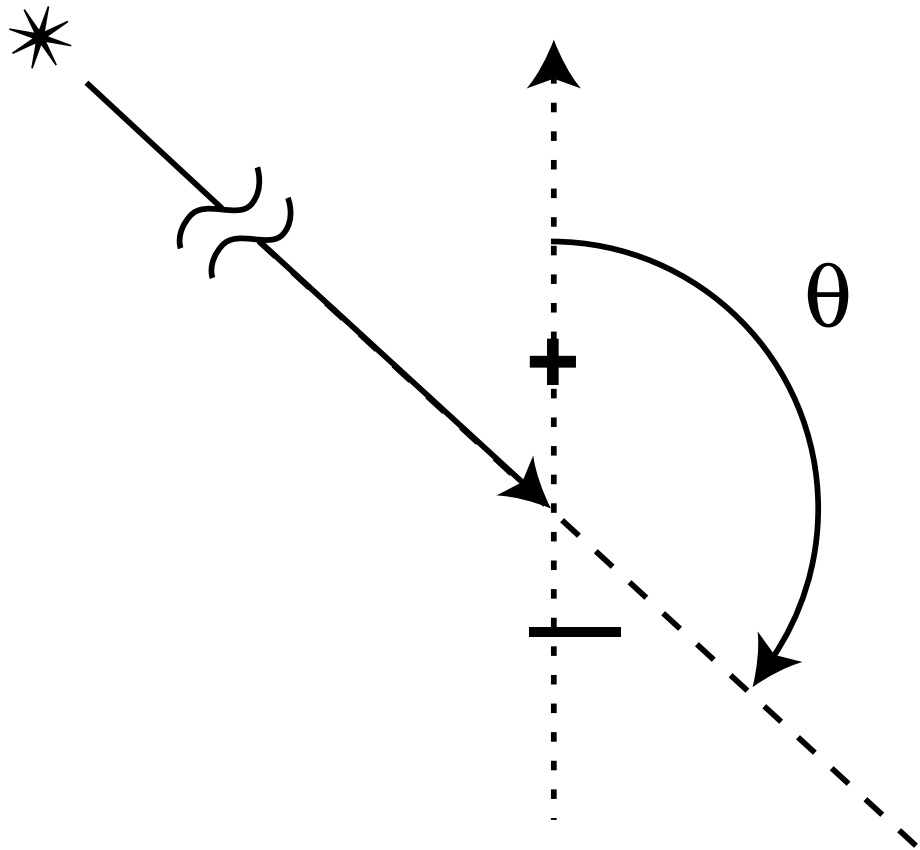}
  \end{center}
  \caption{To illustrate the effectiveness of a correlation analysis 
    compared to a coincidence analysis we apply both to a model
    problem involving two identical detectors of isotropic response.
    The detectors are denoted $+$ and $-$. Their relative separation
    and the parameters describing the direction of an incident plane
    wave signal are shown in this figure.}\label{fig:toyDet}
\end{figure}

Signals are also characterized by their incident direction relative to
the receiver.  Since the two detectors have an isotropic antenna
pattern, the receiver response to radiation incident from different
directions depends only on the angle between the axis defined by the
two detectors and the radiation's propagation direction.  Figure
\ref{fig:toyDet} shows the geometry we use to describe the interaction
of the model receiver with an incident gravitational wave, with the
cosine of this angle by $X_{0}$.

Again for the purpose of illustration, consider a model astrophysical
burst source population whose members have a well-determined waveform
of finite duration.  Assume the sources are standard candles and
radiate isotropically, with radiation waveform two cycles of a
sine wave of known frequency $f_0$.  Denoting the signal arrival time
at the midpoint between the detectors by $T_{0}$ and the signal
amplitude as $A_{0}$, the response of the two detectors to the signal
is
\begin{mathletters}
  \label{eq:DetSrc}
  \begin{equation}
    s_{\pm} :=\left\{
      \begin{array}{ll}
        A_{0}\sin2\pi f_{0}\left[t-\left(T_{0}\pm RX_{0}\right)\right]&
        \text{if $0<f_{0}\left[t - \left(T_{0}\pm RX_{0}\right)\right]<2$}\\
        0&\text{otherwise.}
      \end{array}
    \right.
  \end{equation}
\end{mathletters}
where we have dropped the distinction between the signal and the
receiver response to the signal and assumed that the detector
bandwidth is much greater than the signal bandwidth.

Since $f_{0}$ is known, the receiver response to a signal is fully 
characterized by $A_{0}$, $T_0$ and $X_{0}$.  Acting alone, the $\pm$ 
detector can measure only $A_{0}$ and $T_{\pm}$, where
\begin{equation}
    T_\pm := T_0 \pm RX_{0}.
\end{equation}

\subsection{Likelihood function}\label{sec:Lambda}

Recall that the receiver's correlation sequence is 
\begin{equation}
    \bbox{C}[j-k] = \overline{\bbox{n}[j]\otimes\bbox{n}[k]},
\end{equation}
where, for our two detector receiver, each $\bbox{C}[k]$ is a 
$2\times2$ matrix.  We have assumed that the noise in detector $+$ is 
uncorrelated with that in detector $-$; consequently, each 
$\bbox{C}[k]$ is diagonal and the matrix $\Bbb{T}_{N_{T}}(\bbox{C})$ 
is conveniently re-organized into a $2\times2$ block diagonal matrix:
\begin{mathletters}
  \begin{equation}
    \left|\left|\Bbb{T}_{N_{T}}(\bbox{C})\right|\right| = 
    \text{diag}\left(\bbox{C}_{{}+{}},\bbox{C}_{{}-{}}\right) = 
    \left(
      \begin{array}{cc}
        \bbox{C}_{+}&0\\
        0&\bbox{C}_{{}-{}}
      \end{array}\right),
  \end{equation}
  where $\bbox{C}_{\pm}$ is the correlation matrix for the $\pm$ 
  detector,
  \begin{equation}
    \left|\left|\bbox{C}_{\pm}\right|\right|_{jk} 
    = \overline{n_{\pm}[j]n_{\pm}[k]}.
  \end{equation}  
\end{mathletters}

Expressed in this way, it is apparent that the likelihood function is
separable:
\begin{mathletters}
    \begin{equation}
        \Lambda(\bbox{g}|A,X,T_{0}) = 
        \Lambda({g}_{+}|A,T_{{}+{}})
        \Lambda({g}_{{}-{}}|A,T_{{}-{}})
    \end{equation}
    where 
    \begin{eqnarray}
        \ln\Lambda({g}_{{}\pm{}}|A,T_{\pm}) &=& 
        2<g_{\pm},m(A,T_{\pm})>_{\bbox{C}_{\pm}} 
        -<m(A,T_{\pm},m(A,T_{\pm})>_{\bbox{C}_{\pm}}
    \end{eqnarray}
\end{mathletters}
The $\Lambda({g}_{\pm}|A,T_\pm)$ are exactly the likelihood functions
for the $\pm$ detectors regarded as single, isolated receivers.  This
separation is always possible when the noise in the detectors is
uncorrelated. When the noise in the detectors is correlated the 
likelihood is still separable once the pseudo-detectors are defined 
through as described in sec.\ \ref{sec:preprocess}.

\subsection{Signal detection}\label{sec:FreqTests}

In this section we consider, in the context of our model receiver, two
different ways one might use a pair of detectors to detect a signal
and infer its parameters.  One procedure exploits the notion of
coincidence: if the two detectors identify separately a signal with
sufficiently similar parameters then the receiver is said to have
detected a signal.  The other procedure exploits the notion of
correlation as developed in section \ref{sec:gauss}: if the response
of an array of detectors is consistent with an incident plane-wave,
then the receiver is said to have detected a signal.  For each of
these two tests we determine the detection efficiency as a function of
the false alarm error fraction when the detector noise is Gaussian. 
We find that the test based on correlation, as embodied in the
receiver likelihood function, has a greater detection efficiency 
than the coincidence test for any choice of false alarm fraction.

\subsubsection{Maximum Likelihood Inference}\label{sec:correlation}

The likelihood function $\Lambda(\bbox{g}|\bbox{\theta})$ is the
dimensionless ratio of two probabilities: the probability of making
the observation $\bbox{g}$ if the signal $\bbox{\theta}$ is present
and the probability of making the observation $\bbox{g}$ if no signal
is present.  It is not a probability itself, nor by itself does it
relate directly to a probability on $\bbox{\theta}$.

Even though we can't regard the likelihood as a measure of the
probability that a signal characterized by $\bbox{\theta}$ is present,
we can regard it as a measure of the \textit{plausibility} of that conclusion:
when $\Lambda(\bbox{g}|\bbox{\theta})$ is greater than unity it
signals that the particular observation $\bbox{g}$ is more likely when
the signal characterized by $\bbox{\theta}$ is present than when no
signal is present.  Similarly, if we assume that a signal is present,
then the parameter $\bbox{\theta}$ that maximizes the likelihood
function is the most \textit{plausible} description of the signal.

Together, these observations motivate a test based on the likelihood
function: when the likelihood function maximum
$\Lambda(\bbox{g}|\widehat{\bbox{\theta}})$ exceeds a threshold
$\Lambda_{0}$, then we conclude that a signal is present and take
$\widehat{\bbox{\theta}}$ to be the maximum likelihood estimator, or
MLE, of the detected signal.

To be precise, consider an observation $\bbox{g}$, whose $N$ samples
$\bbox{g}[k]$ are taken at time $t_{k}$.  Denote by
$\bbox{s}(\bbox{\theta})[j]$ the sampled receiver response to a signal
characterized by $\bbox{\theta}$.  Assume that the observation
duration is much longer than the longest signal response (so that we
need not consider signals that begin before or end after the
observation period).  Denote by $\bbox{\theta}_{k}$ the parameter
space of signals whose leading wavefront is incident on the receiver
at time $t_{k}$: in our example, these are just $A$ and $X$.  Finally,
fix a threshold $\rho^{2}_{0}$.  The following procedure produces a
list of detected signals and point estimates of the parameters
describing each:
\begin{enumerate}
\item Evaluate the log-likelihood function
  $\ln\Lambda(\bbox{g}|\bbox{\theta}_{k})$ for signals incident on the
  receiver at the sample times $t_{k}$.
  
\item{}\label{step:ml}At each sample time $t_{k}$, find the signal
  characterization $\widehat{\bbox{\theta}}_{k}$ that maximizes
  $\ln\Lambda(\bbox{g}|\bbox{\theta}_{k})$. Associate with each
  $t_{k}$ and $\widehat{\bbox{\theta}}_{k}$ a S/N
  $\widehat{\rho}_{k}$, given by
  \begin{equation}
    \widehat{\rho}^{2}_{k} = 
    \ln\Lambda(\bbox{g}|\widehat{\bbox{\theta}}_{k}).
  \end{equation}
    
  \item{}\label{step:candidate}Order the triplets
  $\{\widehat{\rho}_{k}, \widehat{\bbox{\theta}}_{k}, t_{k}\}$ with
  respect to $t_{k}$.  Select the subset of triplets where \textit{i)}
  $\rho_{k}$ is greater than the threshold $\rho_{0}$ and \textit{ii)} a
  local maxima; i.e., find the
  $\{\widehat{\rho}_{k'},\widehat{\bbox{\theta}}_{k'}\}$ for which
  \begin{mathletters}
    \begin{eqnarray}
      \rho_0 &<& \widehat{\rho}_{\pm,k'}\\
      \widehat{\rho}_{\pm,k'-1} &\leq& \widehat{\rho}_{\pm,k'}\\
      \widehat{\rho}_{\pm,k'+1} &\leq& \widehat{\rho}_{\pm,k'}.
    \end{eqnarray}
  \end{mathletters}
        
\item{}\label{step:prune}Beginning with the largest
  $\widehat{\rho}_{k}$ in this subset find all other triplets
  $\{\widehat{\rho}_{k'}, \widehat{\bbox{\theta}}_{k'}, t_{k'}\}$ for
  which $|t_{k}-t_{k'}|$ is less than the signal duration.  Discard
  these tripletts.  Repeat with the next largest remaining
  $\widehat{\rho}_{k}$ until the list is exhausted.  What remains is
  the list of \textit{detected signals,} with S/N $\widehat{\rho}_{k}$,
  signal start time $t_{k}$, and characterized by (the point estimate)
  $\widehat{\bbox{\theta}}_{k}$.
\end{enumerate}

Three steps in the maximum likelihood test procedure deserve 
additional discussion: the focus only on bursts starting at the 
discrete sample times $t_{k}$ (step \ref{step:ml}), the formation 
of the intermediate list of consisting of local maxima of the maximum 
of the likelihood function (step \ref{step:candidate}) and the 
pruning of this list to form the final list of detected signals (step
\ref{step:prune}). 

Step \ref{step:ml} focuses attention on signals arriving at the
discrete sample times.  Real signals, however, are not so constrained. 
Nevertheless, if the observation is properly sampled (i.e., sampled
without aliasing), then all of the power in the receiver response is
at frequencies much less than the Nyquist frequency $f_{N}$, which is
half the sample rate.  In that case
$\left<\bbox{m}(\bbox{\theta},t_{0}),
\bbox{m}(\bbox{\theta},t_{0}+\tau)\right>_{\bbox{C}}$, where
$\bbox{m}(\bbox{\theta},t)$ denotes the receiver response to a signal
whose initial wavefront arrives at the network at time $t$, cannot
vary significantly for $|\tau|$ less than several times $1/f_{N}$;
correspondingly, the likelihood will remain peaked about the $t_{k}$
nearest to the actual signal arrival time and the corresponding signal
to noise will differ only slightly from its maximum
value.\footnote{The errors incurred here can be reduced still further
by appropriate interpolation.}

In step \ref{step:candidate}, we select only the local maxima of the
likelihood function as candidate signal events. This reflects the
observation that, in the absence of noise, the likelihood function is
maximized when $\bbox{\theta}$ is equal to the true signal
characterization $\bbox{\theta}_{t}$. 

Even in the absence of noise, however, not all local maxima can be 
identified as distinct signals.  While the likelihood function is 
maximized when $\bbox{\theta}$ is equal to the true signal 
characterization $\bbox{\theta}_{t}$, as $\bbox{\theta}$ differs from 
$\bbox{\theta}_{t}$ the likelihood decreases, but not necessarily 
monotonically.  Even for our simple signal model there are three local 
maxima associated with the likelihood function.  The situation is 
further complicated when, as is the actual case, receiver noise 
distorts the ``noise-free'' likelihood, randomly increasing it for 
some $\bbox{\theta}_{k}$ and decreasing it for others. 

To help distinguish between the global maximum of the likelihood
function and its sidelobes, we make use of our implicit assumption
that real signals are sufficiently rare that the receiver response to
one real signal does not have a significant probability of overlapping
with its response to a second real signal.  Any two local maxima
separated in time by less than the signal duration are then associated
with a single source.  In step \ref{step:prune} we prune the list of
candidate signals (i.e., the local maxima identified in step
\ref{step:candidate}) by identifying clusters of local maxima and
replacing each with its single, strongest member.\footnote{This intrinsically
non-linear step introduces into the analysis a notion of detector
``dead time:'' i.e., the analysis is unable to identify more
than a single signal in any given interval of duration less than the
signal duration.}  The result is a list of events, all above threshold,
in which no two events can have resulted from the same
gravitational-wave signal.

Finally, we justify the use of $\widehat{\bbox{\theta}}$ as the point
estimate of the signal parameters characterizing the detected signal. 
Suppose that a signal characterized by fixed $\bbox{\theta}_{t}$ is
incident on an ensemble of identical receivers.  The corresponding
ensemble of $\widehat{\bbox{\theta}}$ has as its mode
$\bbox{\theta}_{t}$; consequently, a natural estimator for
$\bbox{\theta}_{t}$ is the $\widehat{\bbox{\theta}}$ arising from a
particular observation.  In the maximum likelihood rule, when we
conclude that a signal is present we take $\widehat{\bbox{\theta}}$ as
our point-estimate of the signal parameters.

\subsubsection{Coincidence inference}\label{sec:coincidence}

Much discussed in the context of gravitational-wave data analysis is
an apparently simpler analysis, referred to generally as
``coincidence.''  This test has received its most precise definition
in \cite{jaranowski94a,jaranowski96a} for the particular case of
binary inspiral observations.

In general, coincidence tests involve a complete analysis at each
individual detector, considered in isolation from all other detectors
in the receiver.  The result of these individual analyses is a set of
``candidate-event lists'', one for each detector, which consist of
``detections'' at each detector together with estimators for the
signal start time and other signal parameters that can be determined
from observations in a single detector.  Real gravitational-wave
events should excite the several different detectors in a
self-consistent manner: in particular, the signal start times should
be consistent with the light travel time between the detectors and
other signal parameters should be consistent with a unique source.

The consistency requirement is difficult to pin-down.  For example, in
the case of our own model detector and source, consistency would
appear to require that the signal arrival times are consistent with
the signal propagation time between the detectors and that the
measured signal amplitudes be equal.  Owing to detector noise,
however, the estimated signal amplitudes will only approximate the
actual amplitudes, and similarly for the signal start times and other
parameters.  For signal amplitudes, then, a window of some breadth
must be defined and signal candidates whose amplitudes fall within the
window are assumed to arise from a real signal.  The choice of window,
its implementation and the procedure for combining separate estimates
of common parameters all affect the false alarm and false dismissal
fractions that characterize the test.

The problem is more complicated in the case of an estimated source
location.  Consider a real signal, incident on the detectors from a
direction nearly perpendicular to the axis between them.  Let the
measured start time $T_{\pm}$ on the $\pm$ detector be
\begin{equation}
T_{\pm} = T_{0,\pm}+\epsilon_{\pm},
\end{equation}
where $T_{0,\pm}$ is the actual moment when the signal is incident on
the $\pm$ detector and $\epsilon_{\pm}$ is the error in the estimated
start time owing to detector noise.  The difference in the measured
signal start times is
\begin{eqnarray}
T_{+}-T_{-} &=& T_{0,+}-T_{0,-} + \epsilon_{+}-\epsilon_{-} \nonumber\\
&=& 2RX_{0} + \Delta{\epsilon},
\end{eqnarray}
where $X_{0}$ is the direction cosine describing the radiation's
propagation direction and $\Delta{\epsilon}$ is the difference between
the errors in the measured start times.  When a real signal is
incident along, or nearly along, the axis ($|X_{0}|\simeq1$), then
small errors $|\Delta\epsilon|\ll2R$ can lead to
$|\widehat{X}_{0}|>1$, which we would regard as unphysical and not
representative of a real signal.  On the other hand, when $|X_{0}|$ is
much less than unity, the same errors will leave us with
$|\widehat{X}_{0}|$ still much less than unity, in which case we
accept the coincidence as representing a real signal.  To minimize
this false rejection of real signals we can adopt a window broader
than the light travel time between the detectors for comparing signal
arrival times (taking the estimated arrival direction to be along the
axis if the difference of arrival times would suggest $|X|$ greater
than unity); however, in doing so we also increase the false alarm
rate, reducing the discriminating power of the test.  

The sign of the error also plays an important role: when
$\Delta{\epsilon}$ has the same sign as $X_{0}$ we are more likely to
reject a real signal than when they have opposite signs.  The fraction
of signals rejected can thus depend in a complicated way on the
interaction between the underlying signal parameters, the windows, and
the allowable range of the parameters that characterize the signal. 

In the spirit of \cite{jaranowski94a,jaranowski96a} we define a coincidence
inference procedure for our model receiver:
\begin{enumerate}
\item For each detector considered in isolation, determine the two
  sets of \textit{candidate signals\/} associated with detector $+$ and
  $-$:
  \begin{enumerate}
  \item Evaluate the log-likelihood function
    $\ln\Lambda({g}_{\pm}|\bbox{\theta}_{\pm,k})$ for signals incident
    on detector $\pm$ at the sample times $t_{\pm,k}$.
    
  \item At each sample time $t_{\pm,k}$, find the signal
    characterization $\widehat{\bbox{\theta}}_{\pm,k}$ that maximizes
    $\ln\Lambda({g}_{\pm}|\bbox{\theta}_{\pm,k})$.  Associate with
    each $t_{\pm,k}$ and $\widehat{\bbox{\theta}}_{\pm,k}$ a
    S/N $\widehat{\rho}_{\pm,k}$, given by
    \begin{equation}
      \widehat{\rho}^{2}_{\pm,k} = 
      \ln\Lambda({g}_{\pm}|\widehat{\bbox{\theta}}_{\pm,k}).
    \end{equation}
    The result is a set of associated signal-to-noise,
    parameterizations and signal start times
    $\{\widehat\rho_{\pm,k},\widehat{\bbox{\theta}}_{\pm,k},
    {t}_{\pm,k}\}$.
    
    \item For the list associated with each detector, select the
    subset $\{\widehat\rho_{\pm,k'},\widehat{\bbox{\theta}}_{\pm,k'},
    t_{\pm,k'}\}$ for which
    \begin{mathletters}
      \begin{eqnarray}
        \widehat\rho_{\pm,0} &<& \rho_{\pm,k'}\\
        \widehat\rho_{\pm,k'-1} &\leq& \rho_{\pm,k'}\\
        \widehat\rho_{\pm,k'+1} &\leq& \rho_{\pm,k'}
      \end{eqnarray}
    \end{mathletters}
    where $\widehat\rho_{\pm,0}$ is the signal detection threshold in
    detector $\pm$.
    
  \item{}\label{step:CSelect}Beginning with the largest
    $\widehat{\rho}_{\pm,k}$ in the list of local maxima associated
    with detector $\pm$, find all other $\widehat{\rho}_{\pm,k'}$ for
    which $|t_{k}-t_{k'}|$ is less than the signal duration.  Discard
    these.  Repeat with the next largest remaining
    $\widehat{\rho}_{\pm,k}$ until the list is exhausted.  What
    remains is the list of \textit{candidate signals,} with
    S/N $\widehat{\rho}_{\pm,0}$, signal start time
    $t_{\pm,k}$, and characterized by (the point estimate)
    $\widehat{\bbox{\theta}}_{\pm,k}$.
  \end{enumerate}
    
\item{}\label{step:CPair}Choose the candidate signal list associated
  with detector $+$.  Beginning with the candidate signal of largest
  S/N $\widehat\rho_{+,k}$ in that list, process
  that list in order of decreasing $\widehat\rho_{+,k}$ to create a
  new, \textit{coincident detection event list:\/}
  \begin{enumerate}
  \item Let the current candidate event from the list associated with
    detector $+$ be numbered $k_{+}$.
      
  \item Identify, from the candidate event list associated with
    detector $-$, all candidates whose start times $t_{j_{-}}$ in
    detector $-$ are consistent with the candidate signal arrival time
    $t_{k_{+}}$ in detector $+$: i.e.,
    \begin{equation}
      |t_{k_{+}}-t_{j_{-}}| < 2R.
    \end{equation}
    Impose other consistency requirements, associated with
    $\widehat{\bbox{\theta}}_{+,k_{+}}$ and
    $\widehat{\bbox{\theta}}_{-,j_{-}}$, as are deemed appropriate.
    (In our model receiver/source example we do not impose any other
    consistency requirements.)
      
  \item The result is a list of candidate coincident events in
    detector $-$ associated with the event $k_{+}$ in detector $+$.
    The list may contain zero, one or more than one event:
    \begin{enumerate}
    \item If it contains no events, delete event $k_{+}$ from the list
      of candidate events associated with detector $+$.
        
      \item If it contains exactly one event (say,
      $\{\widehat{\rho}_{-,j_{-}}$,
      $\widehat{\bbox{\theta}}_{-,j_{-}}$, $t_{-,j_{-}}\}$), pair it
      with the event $\{\widehat{\rho}_{+,k_{+}}$,
      $\widehat{\bbox{\theta}}_{+,k_{+}}$, $t_{+,k_{+}}\}$ from the
      list associated with detector $+$ and add the pair to the
      \textit{coincident detection event list.} Delete all events from
      the 
      candidate list associated with detector $+$ whose arrival times
      are so close that they would overlap with event $k_{+}$;
      similarly, delete all events from the candidate list associated
      with detector $-$ whose arrival times are so close that they
      would overlap with event $j_{-}$.  Delete events $k_{+}$ and
      $j_{-}$ from their respective candidate lists.
    
      \item If it contains more than one event, choose the single
      event $j_{-}$ with greatest strength $\widehat{\rho}_{-,j}$,
      pair it with event $k_{+}$ from the candidate list associated
      with detector $+$, and add the pair to the coincident detection
      event list.  Delete all events from the candidate list
      associated with detector $+$ whose arrival times $t_{+,k}$ are
      so close that they would overlap with event $k_{+}$; similarly,
      delete all events from the list associated with detector $-$
      whose arrival times $t_{-,j}$ are so close that they would
      overlap with event $j_{-}$.  Delete events $k_{+}$ and $j_{-}$
      from the lists associated with the respective detectors.
    \end{enumerate}
  \end{enumerate}
\end{enumerate}
The result of applying this procedure to the output of the $\pm$
detectors is a set of paired events, one from each detector.  Each
member of the set involves a pair of signal amplitudes (in this case,
equivalent to S/N) and the best estimate of the signal arrival time at
each detector.  The signal arrival times are, by construction,
consistent with the incidence of a plane wave on the detector pair.

It remains to combine the signal arrival times and amplitudes in each 
pair to determine a single estimate of the signal amplitude, the 
signal arrival time at the mid-point between the detector, and the 
radiation propagation direction. In our model problem, the 
natural estimators for the latter two quantities are
\begin{mathletters}
  \label{eq:combinedEstimators}
  \begin{eqnarray}
    \widehat{T}_{j} &=& \left(t_{+,j} + t_{-,j}\right)/2\\
    \widehat{X}_{j} &=& \left(t_{+,j} - t_{-,j}\right)/2R.\label{eq:coincEstX}
  \end{eqnarray}
\end{mathletters}

The geometry of our model problem suggests no particular procedure for
combining the separate signal amplitude estimates into an overall
estimate for the network.  One procedure that has been recommended
\cite{jaranowski94a,jaranowski96a} is to form the final estimate as
the root mean square of the point estimates:
\begin{mathletters}
\begin{equation}
\widehat{A}^{2}_{j} = {1\over2}\left(
\widehat{A}_{+,j}^{2} +
\widehat{A}_{-,j}^{2}\right),
\end{equation}
where
\begin{equation}
    \widehat{A}_{\pm} := 
    {\rho^{2}_{\pm}\over\left<s_{\pm},s_{\pm}\right>_{C_{\pm}}}
\end{equation}
\end{mathletters}
This prescription will consistently overestimate the amplitude of the 
signal. For any given observation of a signal with amplitude $A_{0}$, 
the estimate in the detector $\pm$ is equal to $A_{0}$ plus a random 
variable:
\begin{equation}
\widehat{A}_{\pm} = A_0 + n_{\pm}. 
\end{equation}
If the detector noise is Gaussian then $n_{\pm}$ is Gaussian. The 
mean square of the point estimates is thus
\begin{equation}
    \widehat{A}^{2} = A_{0}^{2} + A_{0}\left(n_{+}+n_{-}\right) + 
    {1\over2}\left(n_{+}^{2}+n_{-}^{2}\right).
\end{equation}
The mean of $\widehat{A}^{2}$, or of $\widehat{A}$, will thus be
greater than $A_{0}$.  A similar problem will plague any attempt to
form network estimates of parameters from parameters that are
overdetermined by the network (for example, network-wide estimates of
$T$ or $X$ from three or more detectors).

An unbiased estimate for $A_{0}$ is the straightforward 
average of the parameter values, which we adopt here:
\begin{mathletters}
\label{eq:AjCoinc}
\begin{equation}
\widehat{A}_{j} := {1\over2}\left(
\widehat{A}_{+,j} +
\widehat{A}_{-,j}\right),
\end{equation}
where
\begin{equation}
    \widehat{A}_{\pm,k} := 
    {\left<g_{\pm},s_{\pm,\widehat{\bbox{\theta}}_{\pm,k}}\right>
    \over\sqrt{
    \left<
    s_{\pm,\widehat{\bbox{\theta}}_{\pm,k}},
    s_{\pm,\widehat{\bbox{\theta}}_{\pm,k}}
    \right>}}.
\end{equation}
\end{mathletters}

Several aspects of this procedure for detecting a signal coincident in 
two detectors and estimating the parameters characteristic of the 
source deserve special attention: 

\paragraph{Candidate signals.} Each event in a pair identified as a 
coincident detection stands on its own as a detection in its detector 
at the given threshold. 

\paragraph{Estimator bias.} When the noise distribution is Gaussian,
the error in the estimator of the signal arrival time at a particular
detector is also Gaussian.  Consequently it might be thought that the
error in the estimators $\widehat{T}_{j}$ and $\widehat{X}_{j}$ are
also normal, since they arise from the combination of normal errors in
two detectors whose noise is normal, and that the estimators
$\widehat{T}_{j}$ and $\widehat{X}_{j}$ are unbiased.  (This is the
claim is made in \cite{jaranowski94a,jaranowski96a}.)  As discussed
above, however, the errors in the $t_{\pm,j}$ are correlated by the
procedure we used to create the coincident detection list: for
radiation whose propagation direction is nearly aligned with the axis
between the detectors, insisting that $|t_{+,j}-t_{-,j}|$ be less than
$2R$ causes us to favor those signals for which the errors in
$t_{+,j}$ and $t_{-,j}$ are positively correlated.  The estimator
$X_{j}$ in equation \ref{eq:coincEstX} is thus biased to underestimate
the magnitude of the propagation direction cosine; additionally,
signals whose propagation direction cosine is large (i.e.,
signals propagated along or nearly along the axis) have larger false
dismissal fractions than signals propagating normal to the axis.

\paragraph{Signal strength.} In the maximum likelihood test described
above, signal strength is described by a single quantity: the S/N,
which is equal to the log maximum likelihood. This measure of 
signal strength has the desired property that, as the detection 
threshold is increased, weaker signals are no longer considered to be 
detected before stronger signals. 
In the
coincidence test, there are two different ``signal-to-noise'' ratios
--- one for each detector --- and neither, by itself, is sufficient to
determine that a signal is present.  It has been suggested
\cite{jaranowski94a,jaranowski96a} that the ``natural'' signal
strength for coincidence tests is the sum of the amplitude-squared
S/N for the different detectors: in this case
\begin{equation}
\widehat{\rho}_{j}^{2} = \widehat{\rho}_{+,j}^{2} + 
\widehat{\rho}_{-,j}^{2}.
\end{equation}
This definition has the undesirable property that ``stronger''
signals (i.e., those with larger $\widehat{\rho}$) are not
necessarily more likely to be detected than weaker ones.  In
particular, as the detection thresholds are raised at the two
detectors, signals disappear from the coincident detection list
\textit{when the weakest member of the pair 
$\{\widehat{\rho}_{+},\widehat{\rho}_{-}\}$ falls below the threshold
in the $\pm$ detector,} which is not when $\rho^{2}$ falls below
threshold.  If we want signal strength to have the property that, as
the detection threshold is increased, weaker signals disappear from
the detection list before stronger ones then the appropriate measure
of signal strength is the \textit{minimum of $\widehat{\rho}_{+}$ and
$\widehat{\rho}_{-}$:\/}\footnote{When the several detectors in the 
network are not identical, or do not have coincident or isotropic 
antenna patterns, then the criteria that weaker signals are always 
less likely to be detected than stronger ones becomes more difficult 
to determine.}
\begin{equation}
\widehat{\rho} \equiv \min(\widehat{\rho}_{+}, \widehat{\rho}_{-}).
\end{equation}

The detection rule described in this section is not the only such rule
in the spirit of coincidence that can be defined.  Many variations are
possible, corresponding to the many \textit{ad hoc\/} decisions that must
be made, especially in identifying candidate events lists for the
separate detectors and identifying ``consistent'' coincidences.  The
choices made here are among the simplest that lead to a well-defined
procedure for identifying coincident events.

\subsubsection{Gaussian Noise}\label{sec:FExample}

To assess the relative performance of the maximum likelihood and 
coincidence inference rules we use Monte Carlo simulations to 
calculate the false alarm and false dismissal fractions $\alpha$ and 
$\beta$ as well as the distributions of the estimators 
$\widehat{T}_0$, $\widehat{X}$ and $\widehat{A}_0$ for a typical 
signal.

An inference rule's false alarm frequency $\dot{N}_{\alpha}$ is the
limiting frequency of ``signal detection'' when, in fact, no signal is
actually present.  To determine $\dot{N}_{\alpha}$ as a function of
the threshold $\rho_{0}^{2}$ we use a statistical model of the
receiver noise to generate many pseudo-random instances of $\bbox{g}$
representative of receiver noise alone.  The false alarm frequency is
then the average number of ``detections'' per unit time.  A
convenient, dimensionless representation of the false alarm frequency
is the average number of false signals detected per sample
$\bbox{g}[k]$:
\begin{equation}
\alpha \equiv \dot{N}_{\alpha}/f_{s},
\end{equation}
where $f_{s}$ is the sample rate.  We refer to $\alpha$ as the false
alarm fraction; by our procedure $\alpha$ is strictly less than or
equal to unity and can be regarded as the probability of a false
detection on a per-sample basis.

The false dismissal frequency $\beta$ of an inference rule is the
limiting frequency with which the rule reports that no signal is
present when, in fact, a signal is present; thus, $\beta$ is a
function of the signal (or the signal population).  Another way to
think about $\beta$ is as the detection efficiency: $1-\beta$ is the
fraction of actual signals that the detection procedure will identify. 
To find $\beta$ we generate many pseudo-random instances of receiver
noise and add to them a specific signal.  The result is many instances
of $\bbox{g}$ corresponding to observations of that source.  The
inference rule will conclude that \textit{no\/} signal is present in
some fraction of these synthetic observations: that fraction is the
false dismissal fraction.

In the case of the maximum likelihood test, $\alpha$ and $\beta$ are
controlled by adjusting the threshold $\rho_{0}$: as $\rho_{0}$ is
increased, $\alpha$ is decreased.  In the case of the coincidence test
described in section \ref{sec:coincidence}, $\alpha$ and $\beta$ are
controlled by adjusting the two thresholds $\rho_{\pm,0}$.  Since, in
our example, the two detectors are identical, we set these equal to 
the same $\rho_{0}$.  The
false dismissal frequency depends on the distribution of signals in
the signal population; for simplicity, we assume that all signals in
the population have the same unknown amplitude $A_{0}$ and sky
location $X_{0}$, which are given in the first column of table
\ref{tbl:beta}.

(More realistically the amplitude $A$ depends inversely on the
distance to the source, its orientation with respect to the detector,
and other parameters.  Corresponding to the source distribution in
space and the other parameters is a distribution for $A$.  Requiring
that $\alpha$ not exceed a certain value sets a threshold
$\Lambda_{0}$, regardless of this distribution.  The false dismissal
frequency $\beta$, on the other hand, depends on this distribution in
a straightforward way.)

\begin{figure}
  \epsfxsize=\columnwidth
  \begin{center}
    \leavevmode\epsffile{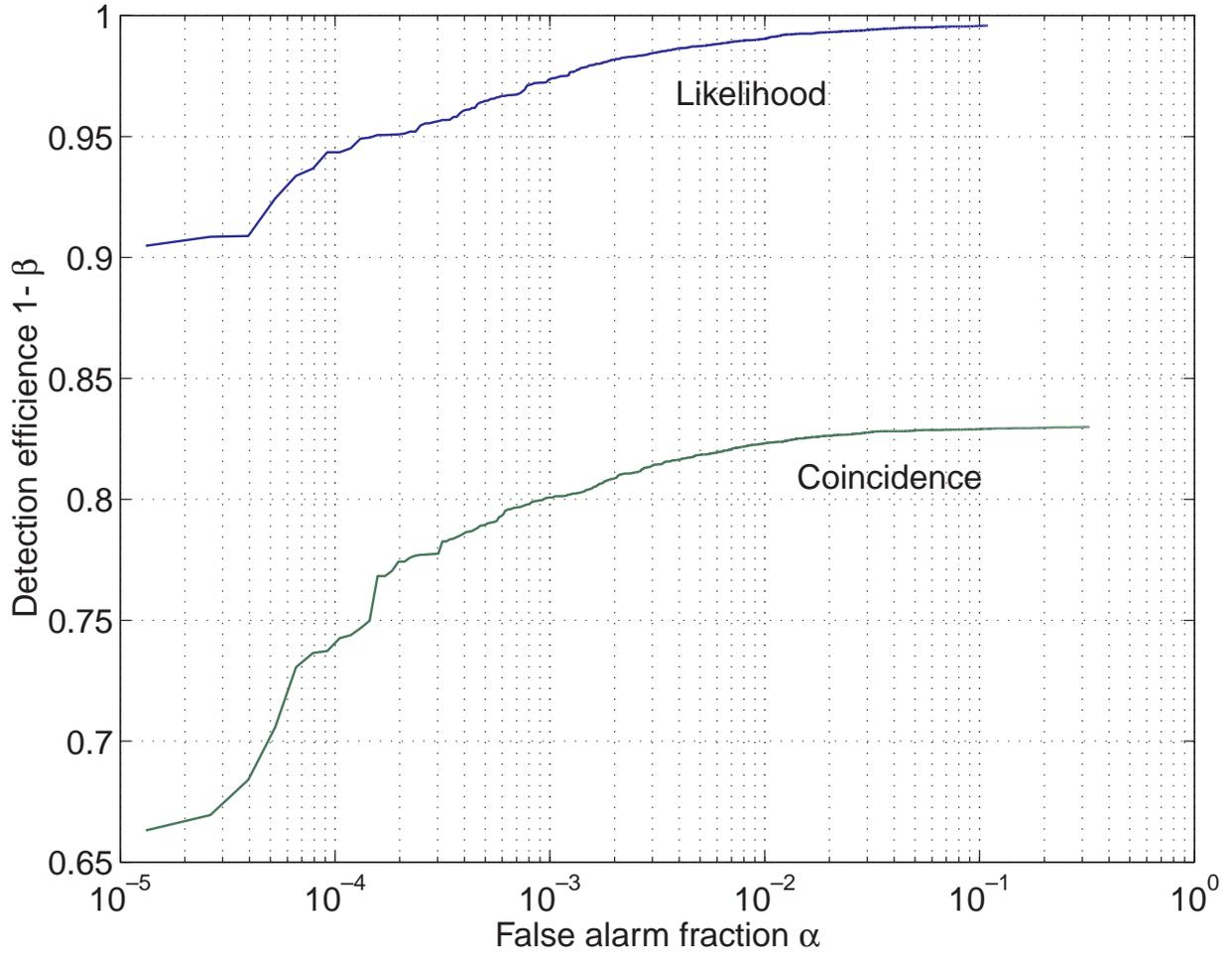}
  \end{center}
    \caption{The false alarm frequency $\alpha$ \textit{vs.}\ the
    detection efficiency ($1-\beta$) for the maximum likelihood and
    coincidence tests in the presence of Gaussian noise.  The
    parameters describing the signal, to which the detection
    efficiency refers, are given in the first column of table
    \protect{\ref{tbl:beta}}.  Note that the performance of the maximum
    likelihood test is everywhere superior to the performance of the
    coincidence test.  The degree of superiority will vary with signal
    strength; however, the relativity performance of the two tests
    will not.  The superior performance of the likelihood based test
    is attributable to the way in which the maximum likelihood test
    internalizes the detector-detector correlations that are present
    when a real signal interacts with the receiver. For more details 
    see section \protect{\ref{sec:FExample}}.}\label{fig:gab}
\end{figure}

Figure \ref{fig:gab} shows $1-\beta$ as a function of $\alpha$ for the
maximum likelihood and coincidence detection procedures when used to
detect signals of this character in our model receiver.  In these Monte
Carlo simulations we count as a false dismissal all signal
identifications (whether by the coincidence or maximum likelihood
test) where the identified start time $\widehat{T}$ differs from the
actual start time by more than the signal duration, or where the
identified sky position $\widehat{X}$ differs from actual sky position
by more than the signal duration divided by the detector separation. 
This condition is necessary if ``correct detections'' by either
rule are to include only those candidate events with non-zero signal
power.  For all $\alpha$, the maximum likelihood test has a
substantially higher detection efficiency $1-\beta$ than the
coincidence test; consequently, its performance is substantially
better than the coincidence test.

The better performance of the maximum likelihood test holds
independent of the actual signal parameters, though it is more
significant for weak signals than for strong ones.  It is, however,
these weak signals --- those just above threshold --- that determine
the overall efficiency of the detector.  For astrophysical burst
sources, which are most likely distributed cosmologically and hence
isotropically, the S/N $\rho$ is inversely proportional to
the source distance; consequently, the number of sources ``brighter''
than the threshold $\rho_0$ is proportional to
$\rho_0^{-3}$. Of these, a fraction $\epsilon$ are ``dimmer'' than
$\rho_\epsilon$, where
\begin{equation}
\rho_\epsilon = \rho_0/\left(1-\epsilon\right)^{1/3}.
\end{equation}
Half of all events whose expected S/N is greater
than $\rho_0$ have S/N less than $1.26\rho_0$. Thus,
if the false dismissal fraction is large when measured for events at
threshold, it will be large when measured over all events. 

Note also in figure \ref{fig:gab} that the false dismissal fraction
for the coincidence test asymptotes to a non-zero value as the false
alarm frequency increases (corresponding to a lower threshold
$\rho_0$): i.e., even at zero-threshold there are false
dismissals.  The asymptote depends on the signal for which the false
dismissal frequency is computed: it is lower for stronger signals and
higher for weaker ones.  The non-zero asymptote for the coincidence
test originates in the process that selects candidate events in each
detector.  In the coincidence test, a false alarm event that occurs
close in time to a real signal event can mask that real event if it
has a higher S/N (cf.\ coincidence test step~\ref{step:CSelect}).  The
false alarm may be sufficiently different than the signal event it
masked that, when an attempt is made to pair it with a candidate event
in the other detectors (cf.\ coincidence test step \ref{step:CPair}),
the test concludes that no signal is present at all; alternatively,
the test may identify a signal at a point in the sky or with a start
time so different from the actual location or start time that the
identification must be regarded as a false alarm and not a signal
detection.

The same mechanism also operates in the maximum likelihood test (cf.\
maximum likelihood test step~\ref{step:prune}); however, that test is
much less sensitive to this effect.  In particular, noise events that
would cause a masking false alarm in the coincidence test do not lead
to a large S/N in the maximum likelihood test since there the S/N is
suppressed when the detector-detector \textit{cross-correlation\/} is not
consistent with a real signal.

The difference in the relative performance of the maximum likelihood
and coincidence tests is directly traceable to the different ways in
which each test requires consistency in the response of the different
detectors: in the maximum likelihood test the \textit{relative\/}
response of the assembled detectors is required to be consistent with
the incidence of a single wavefront on the receiver, while in the
coincidence test the individual response of each detector is
represented in just a few parameters and an \textit{ad hoc\/} consistency
is imposed only on the relative value of these parameters, \textit{none
  of which sample the correlated response of the several detectors in
  the receiver.}
  
\subsubsection{Correlated Noise}\label{sec:correlated}
The example just given is in the context of noise uncorrelated
between the two detectors. How do the coincidence and correlation test
fare when the noise in the several receiver detectors is correlated?

In the context of the coincidence test, correlated noise leads to an
increase in the overall false alarm frequency as noise events leading
to a candidate event in one detector are correlated with noise events
leading to a candidate in the other detector.  No means of
distinguishing between these new false alarms, which arise from the
noise cross-correlation, and correlations arising from signals is
possible in a coincidence test; consequently, the only way that the
correlated noise can be accommodated is by an increase in the
thresholds applied to the output of each detector.  This increases the
false dismissal fraction, leading to an overall worsening of the
test's performance.

The likelihood function, on the other hand, directly accommodates
correlated noise in a precise manner.  In the context of the maximum
likelihood test, noise correlated between the detectors means that the
$\bbox{C}[k]$ are no longer diagonal; however, as we have seen (cf.\ 
sec.\ \ref{sec:fast<a,b>}) this poses no analysis problems in either
principle or practice. By construction, then, the maximum likelihood
test distinguishes inter-detector correlations whose spectrum is
characteristics of a real signal from inter-detector correlations that
are characteristic of correlated detector noise.  Consequently, when
noise is correlated between the receiver's detectors we expect the
maximum likelihood test to perform still better than the coincidence
test.

\subsection{Non-Gaussian Noise}\label{sec:ngdisc}

Equation \ref{eq:LambdaDetSig} describes the likelihood function only
when the receiver noise is Gaussian.  The noise in a real detector
will be, at some level, non-Gaussian and non-stationary: some
fundamental contributions to the noise may be intrinsically
non-Gaussian, some contributions may be intrinsically Gaussian but
appear non-Gaussian in the output owing to non-linearities in the
receiver's response, and some contributions will reflect the
environment that the detector finds itself in.\footnote{We distinguish
  between noise transients, which are generally relatively short
  bursts, and non-stationarity, which we use to mean adiabatic changes
  in the statistical character.} We have already shown that, lacking
detailed knowledge of the higher order moments of the detector noise
distribution, a Normal distribution is the the best approximation to a
noise distribution whose mean and variance are known (cf.\ 
\ref{sec:likeIntro}).  We refer to this as the Gaussian-approximation
likelihood function.

We expect that the maximum likelihood test will still outperform the
coincidence test, since it is still the case that the correlation
analysis described in section~\ref{sec:correlation} is sensitive to
the coherent response of a receiver's component detectors to a real
signal in ways that the coincidence analysis described in
section~\ref{sec:coincidence} is not.  Coincidence tests misidentify
as signals coincident non-Gaussian noise events as readily as they do
coincident Gaussian noise events, while a Gaussian-approximation
likelihood test reject non-Gaussian events that are inconsistent with
an incident plane gravitational wave as easily as they do inconsistent
Gaussian events.  Thus, we expect in general that the detection
efficiency for fixe false alarm fraction will be greater for a test
based on the Gaussian-approximation likelihood test statistic than for
a coincidence test based on the individual detector responses.

To demonstrate this point, we simulate non-Gaussian noise according to
two models --- one strongly leptokurtic and one strongly platykurtic
--- and apply the coincidence and Gaussian-approximation maximum
likelihood tests described in sections \ref{sec:correlation} and
\ref{sec:coincidence} to calculate the relationship between detection
efficiency and false alarm fraction for a fixed signal.  A convenient
model for a stationary non-Gaussian noise process is the
\textit{mixture Gaussian.} A mixture Gaussian distribution has the
form
\begin{mathletters}
\begin{eqnarray}
P\left(x|\{p_{i},\mu_{i},\sigma_{i}, i=1\ldots N\}\right) &=& 
\sum_{i=1}^{N}
p_{i}{\exp\left[{(x-\mu_{i})^{2}/2\sigma_{i}^{2}}\right]\over
\sqrt{2\pi}\sigma_{i}}\\
\sum_{i=1}^{N}p_{i} &=& 1\\
p_{i} &>& 0.
\end{eqnarray}
\end{mathletters}
By appropriate choice of the constants $p_{i}$, $\mu_{i}$ and 
$\sigma_{i}$ a mixture Gaussian can approximate any uncorrelated 
noise distribution through its first $2N$ moments.

\begin{figure}
    \epsfxsize=\columnwidth
   \begin{center}
       \leavevmode\epsffile{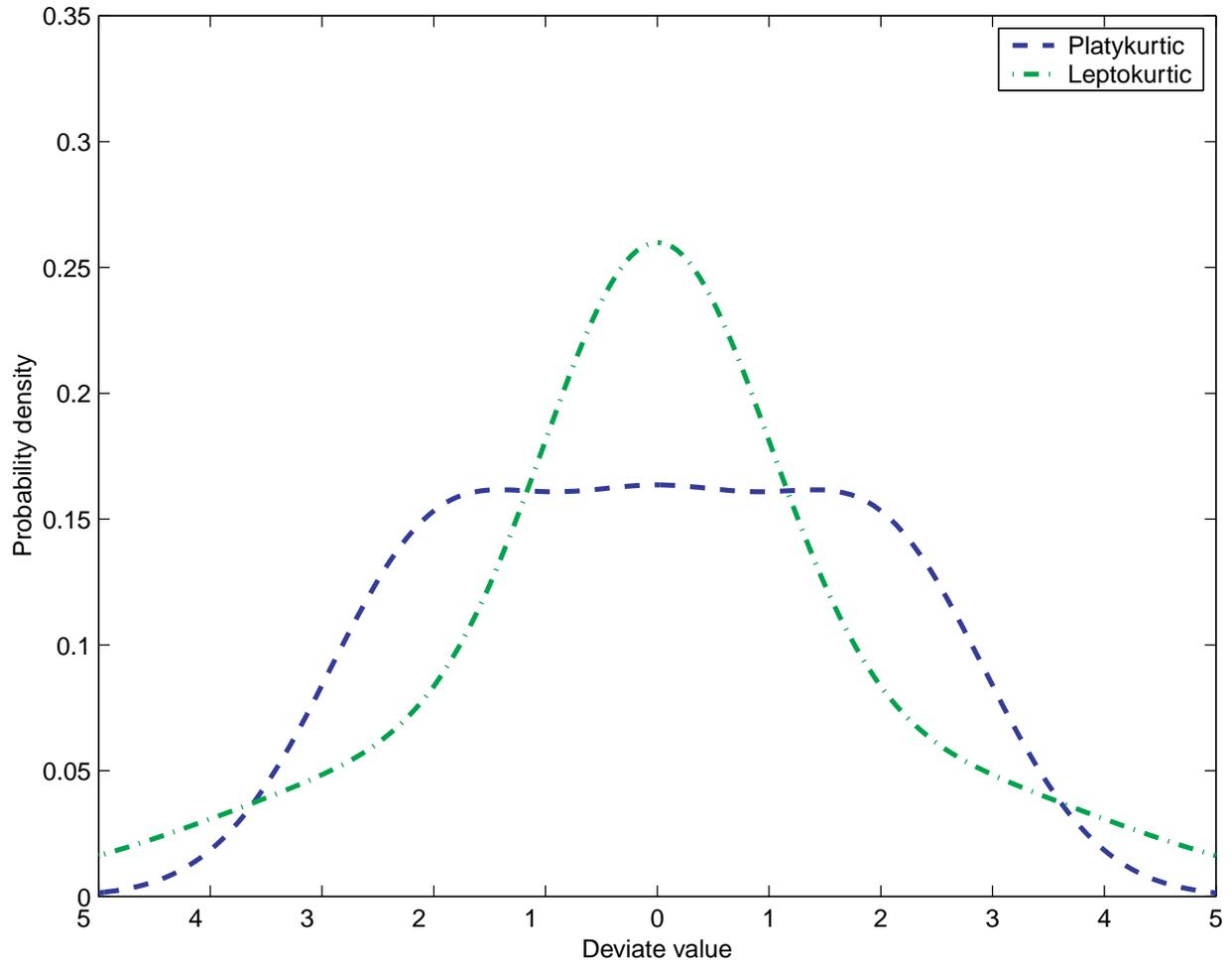}
   \end{center}
    \caption{The probability distribution functions for the
    leptokurtic and platykurtic non-Gaussian noise models used to
    test the relative performance of the coincidence and
    Gaussian-approximation likelihood test.  For more details see
    table \protect{\ref{tbl:mixGaussDist}} and section 
    \protect{\ref{sec:ngdisc}}.}\label{fig:mixGaussDist}
\end{figure}

The two distributions we model here are drawn from the mixture
Gaussian distributions described in table \ref{tbl:mixGaussDist}.  The
corresponding probability distribution functions are shown graphically
in figure \ref{fig:mixGaussDist}.  Note that each is strongly
non-Gaussian, though in different ways.

\begin{figure}
\epsfxsize=\columnwidth
\begin{center}
\leavevmode\epsffile{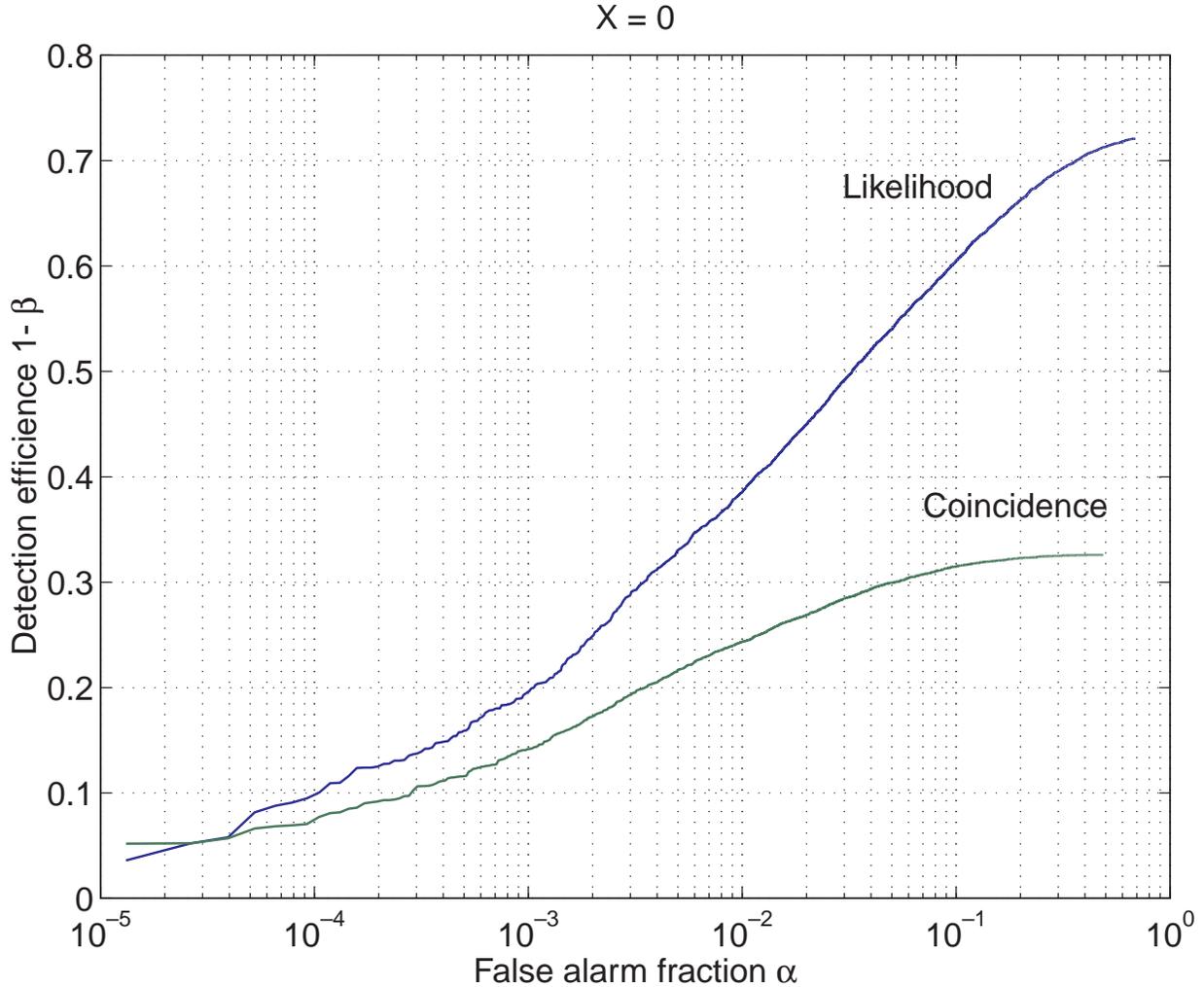}
\end{center}
\caption{The false alarm fraction $\alpha$ \textit{vs.}\ the detection
efficiency ($1-\beta$) for the Gaussian-approximation likelihood and
coincidence tests in the presence of strongly leptokurtic non-Gaussian
noise.  The noise is described by the mixture Gaussian model whose
parameters are given in the first column of table
\protect{\ref{tbl:mixGaussDist}} and the signal used for calculating
the detection efficiency is described in table
\protect{\ref{tbl:beta}}.  Figure \protect{\ref{fig:mixGaussDist}}
shows the noise PDF graphically.  Note that, even though when noise is
substantially non-Gaussian, the Gaussian-approximation likelihood test
has significantly better performance then the coincidence test.  The
superiority is attributable to the fact that real signals are
correlated between the detectors, and the Gaussian approximation
likelihood test, even when the noise is not Gaussian, is still
sensitive to those correlations.  For more discussion see section
\protect{\ref{sec:ngdisc}}.}\label{fig:lepto}
\end{figure}

\begin{figure}
\epsfxsize=\columnwidth
\begin{center}
\leavevmode\epsffile{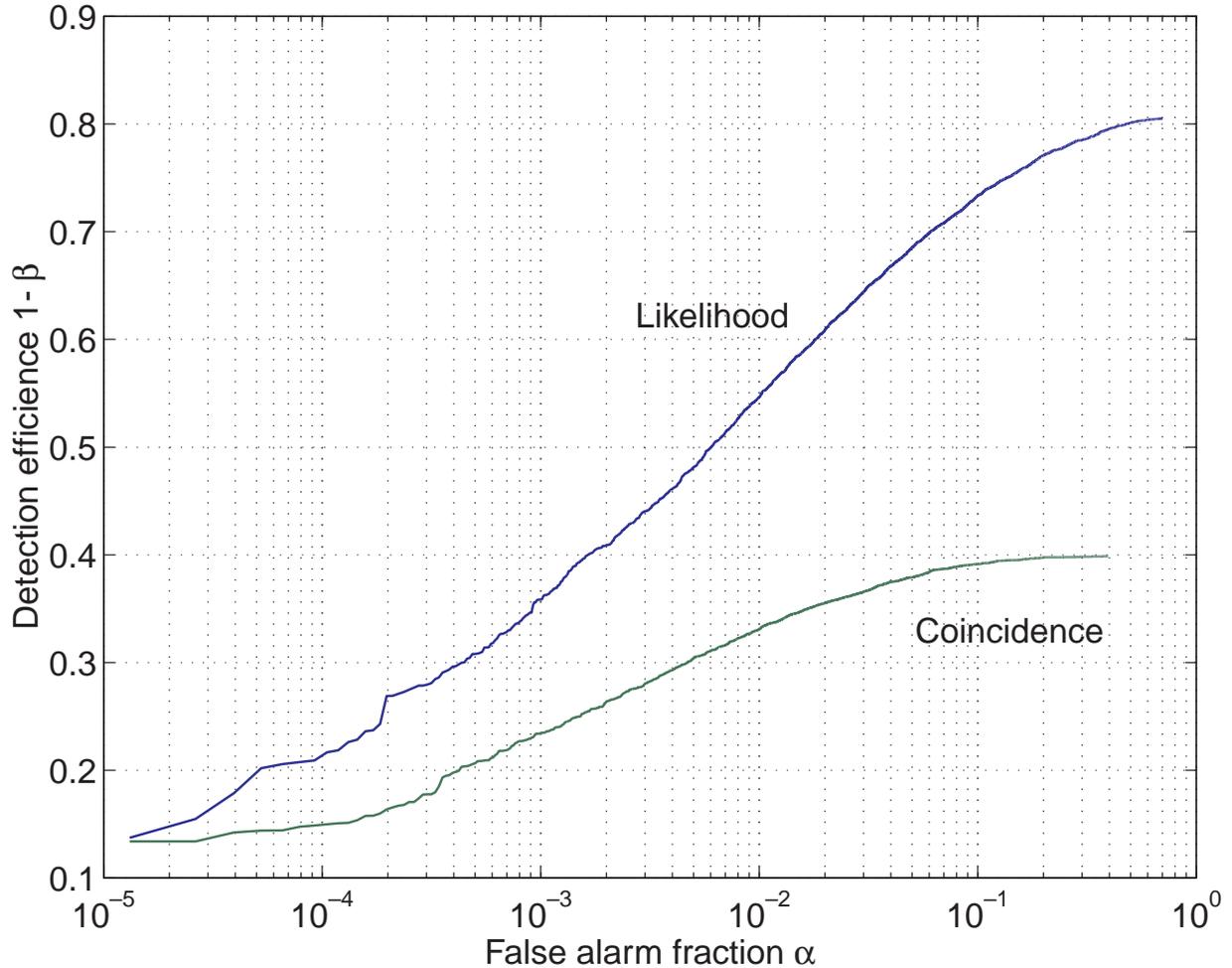}
\end{center}
\caption{The false alarm fraction $\alpha$ \textit{vs.} the detection
efficiency ($1-\beta$) for the Gaussian-approximation likelihood and
coincidence tests in the presence of strongly platykurtic non-Gaussian
noise.  The noise is described by the mixture Gaussian model whose
parameters are given in table
\protect{\ref{tbl:mixGaussDist}} and whose PDF is shown graphically in
figure \protect{\ref{fig:mixGaussDist}}.  For more discussion see
section \protect{\ref{sec:ngdisc}}.}\label{fig:platy}
\end{figure}

Figures \ref{fig:lepto} and \ref{fig:platy} show the detection
efficiency as a function of the false alarm frequency for the
coincidence and Gaussian-approximation maximum likelihood tests for
the leptokurtic distribution and platykurtic distributions,
respectively.  The detection efficiency and false alarm fractions were
determined by Monte Carlo simulations.  The signal parameters used in
the detection efficiency simulations are given in the second and third
columns of table \ref{tbl:beta}.  The conclusion reached earlier ---
that the maximum likelihood test is superior to the coincidence test
--- is \textit{not sensitive\/} to the approximation of Gaussian
noise.  There is no qualitative difference between figures
\ref{fig:lepto}, \ref{fig:platy}, and \ref{fig:gab}, which summarize
the relative characteristics of these two tests when the noise is
strongly non-Gaussian or Gaussian.  In all cases the detection
efficiency of the (Gaussian-approximation) maximum likelihood test is
less than that of the coincidence test at the same false alarm
fraction.  The principal reason for this superior performance is the
same here as in the case of Gaussian noise: the response of two or
more detectors to incident radiation is correlated, and the
Gaussian-approximation maximum likelihood test is sensitive to the
expected inter-detector cross-correlations, reducing the S/N when the
correlations are not consistent with waves from the same source.

\section{Conclusions}\label{sec:conclusions} 

The output of several gravitational wave detectors can be combined, in
a form of aperture synthesis, to form a single, more sensitive
gravitational wave detector.  Here we describe such an analysis, based
on the likelihood function, appropriate to the detection of a burst
gravitational wave source, of known waveform, in a network of
gravitational wave detectors.  This likelihood analysis of the joint
output of several detectors leads to the optimal matched filter for
the output of the multi-detector network. 

The analysis presented here stands in contrast to ``coincidence''
analyses, where the output of each detector is studied separately to
arrive at a list of events, which are then compared between the
detectors to determine if there are any coincidences, which may be
taken to be evidence for gravitational waves.

The critical difference between these two analyses is that the
likelihood is sensitive to the coherent response of the detector
network to the incident signal.  This leads the likelihood based
analysis to have greater discriminating power than a coincidence
analysis, as shown by a greater detection efficiency to false
alarm frequency ratio.

The importance of inter-detector correlations is clearly important 
when looking for a signal; however, it is also important when the 
noise in two or more detectors is correlated. Such correlations are 
naturally accommodated in the optimal filter developed here; however, 
they cannot be naturally accommodated in a coincidence analysis which, 
by its very nature, ignores inter-detector correlations. 

The likelihood function derived here begins with an assumption that
the detector noise is Gaussian-stationary; however, the results
obtained are much more general.  We show that treating non-Gaussian as
if it were Gaussian is, in a very well-defined sense, the most
appropriate choice if the only available characterization of the noise
is through its mean and correlation function.  While it is sometimes
claimed \cite{creighton99a} that coincidence is the only reasonable
test when the noise is non-Gaussian, we show in a series of numerical
simulations that the even when the noise is strongly non-Gaussian the
likelihood test, treating the noise as if it were Gaussian,
outperforms the coincidence test as measured by the ratio of detection
efficiency to false alarm fraction.

A naive estimate of the computational cost of computing the 
matched filter for a network of detectors might suggest that the cost 
is proportional to the cube of the length of the time series and the 
number of detectors. If the calculation is properly organized, 
however, the cost is seen to be strictly proportional to the duration 
of the signal being sought and no more than the square of the number 
of detectors in the network. 

The work described here can be considered aperture synthesis
specialized to the problem of searching for signals of finite duration
whose waveform is known.  The problem of searching for stochastic
signals remains to be studied.  True aperture synthesis --- searching
for fringes in the interfered output of several gravitational wave
detectors --- may well be the most sensitive means of searching for 
the truly unanticipated source and is a particularly promising direction 
for future research. 

\acknowledgements

I am especially grateful to Joseph Romano for a very careful reading
of early versions of this manuscript and to Albert Lazzarini for
pointing out that the likelihood can always be made to separate.

This work was supported by grants from the Alfred P. Sloan Foundation
and the National Science Foundation (PHY 93-08728, PHY 98-00111 and
PHY 99-96213).


\begin{thebibliography}{10}

\bibitem{luck00a}
H. L{\"u}ck {\it et~al.},  in {\em Gravitational Waves}, No.~523 in {\em AIP
  Conference Proceedings}, edited by S. Meshkov (American Institute of Physics,
  Melville, New York, 2000), pp.\ 119--127, proceedings of the Third Edoardo
  Amaldi Conference.
See Ref.\ \cite{meshkov00a}.

\bibitem{coles00a}
M. Coles,  in {\em Gravitational Waves}, No.~523 in {\em AIP Conference
  Proceedings}, edited by S. Meshkov (American Institute of Physics, Melville,
  New York, 2000), proceedings of the Third Edoardo Amaldi Conference.
See Ref.\ \cite{meshkov00a}.

\bibitem{marion00a}
F. Marion,  in {\em Gravitational Waves}, No.~523 in {\em AIP Conference
  Proceedings}, edited by S. Meshkov (American Institute of Physics, Melville,
  New York, 2000), pp.\ 110--118, proceedings of the Third Edoardo Amaldi
  Conference.
See Ref.\ \cite{meshkov00a}.

\bibitem{conti00a}
L. Conti {\it et~al.},  in {\em Gravitational Waves}, No.~523 in {\em AIP
  Conference Proceedings}, edited by S. Meshkov (American Institute of Physics,
  Melville, New York, 2000), proceedings of the Third Edoardo Amaldi
  Conference.
See Ref.\ \cite{meshkov00a}.

\bibitem{dewaard00a}
A. de~Waard and G. Frossati,  in {\em Gravitational Waves}, No.~523 in {\em AIP
  Conference Proceedings}, edited by S. Meshkov (American Institute of Physics,
  Melville, New York, 2000), proceedings of the Third Edoardo Amaldi
  Conference.
See Ref.\ \cite{meshkov00a}.

\bibitem{blair00a}
D. Blair,  in {\em Gravitational Waves}, No.~523 in {\em AIP Conference
  Proceedings}, edited by S. Meshkov (American Institute of Physics, Melville,
  New York, 2000), proceedings of the Third Edoardo Amaldi Conference.
See Ref.\ \cite{meshkov00a}.

\bibitem{finn92a}
L.~S. Finn, Phys. Rev. D {\bf 46},  5236  (1992).

\bibitem{finn93a}
L.~S. Finn and D.~F. Chernoff, Phys. Rev. D {\bf 47},  2198  (1993).

\bibitem{cutler94a}
C. Cutler and {\'E}.~{\'E}. Flanagan, Phys. Rev. D {\bf 49},  2658  (1994).

\bibitem{jaranowski94a}
P. Jaranowski and A. Krolak, Phys. Rev. D {\bf 49},  1723  (1994).

\bibitem{jaranowski96a}
P. Jaranowski, K.~D. Kokkotas, A. Kr{\'o}lak, and G. Tsegas, Class. Quantum
  Grav. {\bf 13},  1279  (1996).

\bibitem{creighton99a}
J. Creighton, Phys. Rev. D {\bf 60},  021101  (1999).

\bibitem{bose99a}
S. Bose, S.~V. Dhurandhar, and A. Pai, Pramana {\bf 53},  1125  (1999).

\bibitem{bose00a}
S. Bose, A. Pai, and S. Dhurandhar, Int.\ J.\ Mod.\ Phys.\ D {\bf 9},  325
  (2000).

\bibitem{oppenheim89a}
A.~V. Oppenheim and R.~W. Schafer, {\em Discrete-Time Signal Processing}, {\em
  Prentice Hall signal processing series} (Prentice-Hall, Englewood Cliffs, New
  Jersey, 1989).

\bibitem{briggs95a}
W.~L. Briggs and V.~E. Henson, {\em The {DFT}: An owner's manual for the
  {D}iscrete {F}ourier {T}ransform} (SIAM, Philadelphia, 1995).

\bibitem{tsubono95a}
K. Tsubono,  in {\em First Edoardo Amaldi Conference on Gravitational Wave
  Experiments}, edited by E. Coccia, G. Pizzella, and F. Ronga (World
  Scientific, Singapore, 1995), p.\ 112.

\bibitem{hamilton97a}
W.~O. Hamilton {\it et~al.},  in {\em Omnidirectional Gravitational Radiation
  Observatory}, edited by W.~F.~V. Jr., O.~D. Aguiar, and N.~S. Magalhaes
  (World Scientific, Singapore, 1997), pp.\ 19--26.

\bibitem{cerdonio97a}
M. Cerdonio {\it et~al.}, Class. Quantum Grav. {\bf 14},  1491  (1997).

\bibitem{astone93a}
P. Astone {\it et~al.}, Phys. Rev. D {\bf 47},  362  (1993).

\bibitem{astone97b}
P. Astone {\it et~al.}, Astroparticle Physics {\bf 7},  231  (1997).

\bibitem{feldman98a}
G.~J. Feldman and R.~D. Cousins, Phys. Rev. D {\bf 57},  3873  (1998).

\bibitem{giunti98b}
C. Giunti, Phys. Rev. D {\bf 59},  053001  (1999).

\bibitem{giunti98a}
C. Giunti, Phys. Rev. D {\bf 59},  113009  (1999).

\bibitem{roe99a}
B.~P. Roe and M.~B. Woodroofe, Phys. Rev. D {\bf 60},  053009  (1999).

\bibitem{poor94a}
H.~V. Poor, {\em An Introduction to Signal Detection and Estimation}, {\em
  Springer texts in electrical engineering}, 2nd  ed. (Springer-Verlag, New
  York, 1994).

\bibitem{ljung99a}
L. Ljung, {\em System Identification: Theory for the User}, {\em Information
  and system sciences}, 2nd  ed. (Prentice Hall, Upper Saddle River, New
  Jersey, 1999).

\bibitem{meshkov00a}
{\em Gravitational Waves}, No.~523 in {\em AIP Conference Proceedings}, edited
  by S. Meshkov (American Institute of Physics, Melville, New York, 2000),
  proceedings of the Third Edoardo Amaldi Conference.

\end{thebibliography}

\begin{table}
\begin{tabular}{lcccc}
&\multicolumn{2}{c}{Functions}&\multicolumn{2}{c}{Sequences}\\
&Time Domain&Frequency Domain&Time Domain&Frequency Domain\\
Scalar&$a(t)$&$\widetilde{a}(f)$&${a}[k]$&$\widetilde{a}[k]$\\
Vector&$\bbox{a}(t)$&$\widetilde{\bbox{a}}(f)$&${\bbox{a}}[k]$&
$\widetilde{\bbox{a}}[k]$\\
Matrix&$\bbox{A}(t)$&$\widetilde{\bbox{A}}(f)$&${\bbox{A}}[k]$&
$\widetilde{\bbox{A}}[k]$
\end{tabular}
\caption{Nomenclature.}\label{tbl:nomenclature}
\end{table}

\begin{table}
\begin{tabular}{lc}
Detector separation $(2R)$&$25/f_0$\\
Sampling frequency&$4f_0$\\
Detector noise PSD ($S_{\pm}$)&$1/f_0$\\
Observation duration ($T_D$)&$100/f_0$
\end{tabular}
\caption{Parameters describing the toy receiver and signal duration
used for examples in this paper. All parameters are given in units of
$f_0$, the reciprocal signal duration.}\label{tbl:toyDet}
\end{table}

\begin{table}
\begin{tabular}{llll}
    &Gaussian&Leptokurtic&Platykurtic\\
    \hline
    $A_{0}$&2.5&3.5&3.0\\
    $X_{0}$&0.0&0.0&0.0
\end{tabular}
\caption{Parameters describing the signal used in the example
calculation of the detection efficiency for the Gaussian, platykurtic
and leptokurtic noise examples.}\label{tbl:beta}
\end{table}

\begin{table}
\begin{tabular}{lll}
    &Leptokurtic&Platykurtic\\
    $p$&$(1/2,1/4,1/4)$&$(61/192,131/384,131/384)$\\
    $\mu$&$(0,2,-2)$&$(0,2,-2)$\\
    $\sigma$&$(1,2,2)$&$(1,1,1)$\\
    \hline
    Mean:&0&0\\
    Std. Dev.:&2.1213&1.9311
\end{tabular}
\caption{Parameters describing the two mixture Gaussian models used to
explore the maximum likelihood and coincidence test performance
non-Gaussian noise in section \protect{\ref{sec:ngdisc}}.  Also shown
are the mean and standard deviation of the distributions.  The
corresponding PDFs are shown graphically in figure
\protect{\ref{fig:mixGaussDist}}.}\label{tbl:mixGaussDist}
\end{table}


\end{document}